\documentclass[manuscript]{aastex62}

\usepackage{amsmath}
\usepackage{tikz}
\usepackage{multirow}
\usepackage{soul}
\setulcolor{red}

\newcommand{\bq}{\begin{equation}}
\newcommand{\eq}{\end{equation}}

\newcommand{\nn}{\nonumber}

\received{}
\revised{}
\accepted{}

\submitjournal{}

\shorttitle{}
\shortauthors{}

\begin{document}

\title{Constraining Screened Modified Gravity by Space-borne Gravitational-wave Detectors}

\correspondingauthor{Rui Niu, Xing Zhang, Wen Zhao}
\email{nrui@mail.ustc.edu.cn, starzhx@ustc.edu.cn, wzhao7@ustc.edu.cn}

\author{Rui Niu}
\affiliation{CAS Key Laboratory for Researches in Galaxies and Cosmology, Department of Astronomy, University of Science and Technology of China, Chinese Academy of Sciences, Hefei, Anhui 230026, China}
\affiliation{School of Astronomy and Space Science, University of Science and Technology of China, Hefei 230026, China}

\author{Xing Zhang}
\affiliation{CAS Key Laboratory for Researches in Galaxies and Cosmology, Department of Astronomy, University of Science and Technology of China, Chinese Academy of Sciences, Hefei, Anhui 230026, China}
\affiliation{School of Astronomy and Space Science, University of Science and Technology of China, Hefei 230026, China}

\author{Tan Liu}
\affiliation{CAS Key Laboratory for Researches in Galaxies and Cosmology, Department of Astronomy, University of Science and Technology of China, Chinese Academy of Sciences, Hefei, Anhui 230026, China}
\affiliation{School of Astronomy and Space Science, University of Science and Technology of China, Hefei 230026, China}

\author{Jiming Yu}
\affiliation{CAS Key Laboratory for Researches in Galaxies and Cosmology, Department of Astronomy, University of Science and Technology of China, Chinese Academy of Sciences, Hefei, Anhui 230026, China}
\affiliation{School of Astronomy and Space Science, University of Science and Technology of China, Hefei 230026, China}

\author{Bo Wang}
\affiliation{CAS Key Laboratory for Researches in Galaxies and Cosmology, Department of Astronomy, University of Science and Technology of China, Chinese Academy of Sciences, Hefei, Anhui 230026, China}
\affiliation{School of Astronomy and Space Science, University of Science and Technology of China, Hefei 230026, China}

\author{Wen Zhao}
\affiliation{CAS Key Laboratory for Researches in Galaxies and Cosmology, Department of Astronomy, University of Science and Technology of China, Chinese Academy of Sciences, Hefei, Anhui 230026, China}
\affiliation{School of Astronomy and Space Science, University of Science and Technology of China, Hefei 230026, China}

\begin{abstract}

The screened modified gravity (SMG) is a unified theoretical framework, which describes the scalar-tensor gravity with screening mechanism. Based on the gravitational-wave (GW) waveform derived in our previous work \citep{liu2018waveforms}, in this article we investigate the potential constraints on SMG theory through the GW observation of the future space-borne GW detectors, including LISA, TianQin and Taiji.
We find that, for the EMRIs consisting of a massive black hole and a neutron star, if the EMRIs are at Virgo cluster, the GW signals can be detected by the detectors at quite high significant level, and the screened parameter $\epsilon_{\rm NS}$ can be constrained at about $\mathcal{O}(10^{-5})$, which is more than one order of magnitude tighter than the potential constraint given by ground-based Einstein telescope. However, for the EMRIs consisting of a massive black hole and a white dwarf, it is more difficult to be detected than the previous case.
For the specific SMG models, including chameleon, symmetron and dilaton, we find these constraints are complementary with that from Cassini experiment, but weaker than those from lunar laser ranging observations and binary pulsars, due to the strong gravitational potentials on the surface of neutron stars. By analyzing the deviation of GW waveform in SMG from that in general relativity, as anticipated, we find the dominant contribution of the SMG constraining comes from the correction terms in the GW phases, rather than the extra polarization modes or the correction terms in the GW amplitudes.

\end{abstract}



\section{Introduction} \label{sec_intro}

General relativity (GR) is always considered as the most successful theory of gravity. However, various difficulties of this theory are also well known. For instance, in the theoretical side, GR has the singularity and quantization problems \citep{dewitt1967quantum,kiefer2007quantum}. In the experimental side, all the observations in cosmological scale indicate the existence of so-called dark matter and dark energy, which might hint the invalidity of GR in this scale \citep{sahni2004dark,cline2013sources}. For these reasons, since it was proposed by Einstein in 1915, a large number of experimental tests have been performed on various scales, from submillimeter-scale tests in the laboratory to the tests at solar system and cosmological scales 
\citep{hoyle2001submillimeter, adelberger2001new, will2014confrontation, Jain2010Cosmological, koyama2016Cosmological, Burrage_2015, Bertoldi2019AEDGE, PhysRevLett.123.061102}. Unfortunately, most of these efforts have focused on the gravitational effects in weak fields. Since the observable gravitational-wave (GW) signals can only be generated in the strong gravitational fields, and are nearly freely propagating in the spacetime once generated \citep{maggiore2008gravitational}, it provides an excellent opportunity to experimentally test the theory of gravity in the strong-field regime \citep{Abbott2016Tests, ligo2019tests, abbott2019tests}. Recently, with the discovery of compact binary coalescence GW signals, by aLIGO and aVirgo collaborations \citep{ligo_1, ligo_2, ligo_3, ligo_4, ligo_5, ligo_6, ligo_7, ligo_8}, testing GR in the strong gravitational fields becomes one of the key issues in the GW astronomy \citep{sathyaprakash2019extreme, KOSTELECKY2016510, miller2019new}.

The testing of GR by the GW observations is to compare the predictions of GW signals in GR and those in the alternative theories, and constrain the differences of them by observations. Therefore, the choice of typical alternative gravitational theory, and the calculation of GW waveforms in the theory is the crucial role \citep{Berti2018, Yunes2013}. A natural alternative to GR is the scalar-tensor theory, which invokes a conformal coupling between matter and an underlying scalar field (see for instance the Brans-Dicke gravity \citep{Brans1961}), besides the standard spacetime metric tensor. The coupling between scalar field and matter leads to the scalar force (fifth force), and the tight experimental constraints \citep{Williams_2012, ADELBERGER2009} require that the fifth force must be screened in high density environments. In the series of our previous works
\citep{PhysRevD.93.124003,zhang2017gravitational,liu2018waveforms,PhysRevD.100.024038,Zhang2019Constraints, Zhang_2019Angular},
we have studied the general scalar-tensor gravity with the screening mechanisms, which can suppress the fifth force in dense regions and allow theories to evade the solar system and laboratory tests, in a unified theoretical framework called screened modified gravity (SMG). In this framework, the chameleon, symmetron, dilaton and $f(R)$ models in the literature are the specific cases of this theory. We have calculated the parametrized post-Newtonian (PPN) parameters \citep{PhysRevD.93.124003}, the post-Keplerian (PK) parameters \citep{Zhang2019Constraints}, the effective cosmological constant \citep{PhysRevD.93.124003}, the effective gravitational constant \citep{PhysRevD.93.124003}, the change in the orbital period of the binary system caused the gravitational radiations \citep{zhang2017gravitational}. Based on these results, we have derived the constraints on the model parameters by considering the observations in solar system, cosmological scale, the binary pulsars and lunar laser ranging measurements \citep{Zhang2019Constraints, PhysRevD.100.024038, zhang2017gravitational, PhysRevD.93.124003}. In addition, in \citep{liu2018waveforms}, we calculated in details the GW waveforms, produced by the compact binary coalescences during the inspiralling stage, and derived the deviations from that in GR, which are partly quantified by the parametrized post-Einsteinian (PPE) parameters. Utilizing these results, we also obtained the potential constraints on the theory by the future ground-based Einstein telescope.

In addition to the ground-based GW detectors, the space-borne detectors are also proposed. In the near future, the mission, including LISA, Taiji and TianQin, will be launched around 2030s \citep{danzmann2016laser, luo2016tianqin, taiji_2017}. Due to the large arm-lengths of these detectors, the sensitive frequency ranges become $(10^{-4},10^{0})$ Hz, lower than those of ground-based detectors, and the extreme-mass-ratio inspirals (EMRIs) are the important GW sources \citep{Amaro_Seoane_2007, LISA2017EMRIs}.
The event rate of the EMRIs is difficult to estimate because it depends on factors that are poorly constrained by observation. According to the estimations of \citep{LISA2017EMRIs}, at least a few EMRIs per year can be detected by LISA irrespective of the astrophysical model. For the most optimistic astrophysical assumptions, this number can reach a few thousands per year.
EMRI normally consists of a stellar compact object, such as a white dwarf (WD), neutron star (NS), or stellar-mass black hole (BH), and a massive BH, which is an excellent source for the test of gravity \citep{Barack2007using, gair2013testing}. In the previous works
\citep{Will2004testing,PhysRevD.65.042002},
the authors have investigated the constraints on Brans-Dicke gravity, massive gravity, etc, assuming the GW signals of BH-NS binaries observed by LISA mission. Similarly, in this article, we will study the constraints of SMG theory by the GW signals produced by the BH-WD, and BH-NS binaries. In our discussion, we will consider both the LISA and TianQin missions. Taiji is similar to LISA \citep{wu2019analytical}, so we suspect the potential constraint from Taiji is also similar to that from LISA. In the calculation, we consider three different cases for the detection. In case one, we constrain the SMG by detecting the extra GW modes. In case two, we constrain the theory by Fisher information matrix analysis, but consider only the restricted GW waveforms, and in case three, we do the same analysis but including the higher order amplitude corrections in the templates. In comparison with the results in these cases, we investigate the contributions of extra polarization modes, and the higher order amplitude corrections in the model constraints.

This paper is organized as follows.
In the section \ref{sec_waveform}, a brief introduction of screened modified gravity are presented, and the Fourier transform of the GW waveforms in SMG is rewritten for conveniently referring to. Two aspects of detectors' information, the noise curves and the antenna pattern functions, which are relevant to our analysis, are introduced in the section \ref{sec_detectors}. In the section \ref{sec_constr}, the method employed in this work and the process used to get the constraints are showed in detail. The results are presented and discussed in the section \ref{sec_result}, where we compare the constraints given by the forecasts of future space-borne detectors with the constraints obtained by the currently experiments in the three specific SMG models. The full waveform of 2.5PN in amplitude 3.5PN in phase with the corrections concerning with the SMG, and the process to derive the antenna pattern functions are given in the Appendixes \ref{app_pn_waveform}, \ref{app_pattern_f}.

Throughout this paper we adopt the units where $c=\hbar=1$. The reduced Planck mass is $M_{\rm Pl} = \sqrt{1/(8\pi G)}$, where $G$ denotes the Newtonian gravitational constant.
{{Since in this article, we consider only the GW sources in the very low redshift range, the redshifts are not explicitly expressed in formulae of this paper. The distance parameter denotes the luminosity distance, and the chirp mass and total mass in this paper denote the directly measured values in detectors' frame.}}

\section{Gravitational waveforms in screened modified gravity} \label{sec_waveform}

The GW waveforms of binaries with circular orbits in general SMG have been calculated in the previous work \citep{ zhang2017gravitational,liu2018waveforms}. In this section, we will present a brief introduction of screened modified gravity and rewrite the formulae of waveforms in the SMG.
The action of a general scalar-tensor theory in the Einstein frame is given by
\bq
S = \int d^4x \sqrt{-g} \left[\frac{M^2_{\rm Pl}}{2}R - \frac12(\nabla\phi)^2 - V(\phi)\right] +S_m\left[A^2(\phi)g_{\mu\nu}, \psi_m\right],
\eq
where $g_{\mu\nu}$ is the matric in the Einstein frame, $g$ is the determinant of the matric, $R$ is the Ricci scalar, $\phi$ is the scalar field, $\psi_m$ is matter field. Here, $V(\phi)$ is a bare potential which characterizes the scalar self-interaction, and $A(\phi)$ denotes a conformal coupling function representing the interaction between scalar field and matter field.
In scalar-tensor theory, the scalar field can affect the effective mass of a compact object. As suggested by \citep{Eardley1975Observable}, the matter action takes the form of
\bq
S_m=-\sum_{a}\int m_a(\phi)d\tau_a,
\eq
where the constant inertial mass of the compact objects are substituted by a function of the scalar field.
The field equations can be got by the variation of the action with respect to $g_{\mu\nu}$ and $\phi$,
\bq
G_{\mu\nu}=8 \pi G(T_{\mu\nu}+T^{\phi}_{\mu\nu}),
\eq
\bq
\nabla_\mu\nabla^\mu\phi = \frac{\partial}{\partial\phi}(V(\phi)-T),
\eq
where $T_{\mu\nu}$ and $T^{\phi}_{\mu\nu}$ are the energy-momentum tensor of matter field and scalar field respectively, and $T$ is the trace of $T_{\mu\nu}$.
The behavior of the scalar field is controlled by both $V(\phi)$ and $T$, by which we define the effective potential
\bq
V_{\rm eff} = V(\phi) -T.
\eq
As shown in the reference \citep{PhysRevD.93.124003}, in the SMG, the effective potential can be rewritten as
\bq
V_{\rm eff} = V(\phi)+\rho A(\phi),
\eq
where $\rho$ is the conserved energy density in the Einstein frame.
In the wave zone, the metric and the scalar field can be expanded around the flat background $\eta_{\mu\nu}$ and the scalar background (the vacuum expectation value (VEV) of the scalar field) $\phi_{\rm VEV}$,
\bq
g_{\mu\nu} = \eta_{\mu\nu} + h_{\mu\nu}, \qquad \phi=\phi_{\rm VEV} + \delta\phi.
\eq
The bare potential $V(\phi)$ and the coupling function $A(\phi)$ can be expanded as
\bq
\begin{aligned}
V(\phi) &= V_{\rm VEV} + V_1\delta\phi + V_2\delta\phi^2 + V_3\delta\phi^3 + \mathcal{O}(\delta\phi^4), \\
A(\phi) &= A_{\rm VEV} + A_1\delta\phi + A_2\delta\phi^2 + A_3\delta\phi^3 + \mathcal{O}(\delta\phi^4).
\end{aligned}
\eq
The effective mass of the scalar field is given by
\bq
m^2_s \equiv \frac{{\rm d}^2V_{\rm eff}}{{\rm d}\phi^2} \Big |_{\phi_{\rm VEV}}=2\left(V_2+\rho_b A_2\right),
\eq
where $\rho_b$ is the background matter density. We can find that the effective mass of the scalar field depends on the ambient matter density. In the conditions, such as the solar system, the matter density is high, the mass of the scalar field is large so that the range of the force corresponding to the scalar field is too short to have detectable effects. By this way, the effects of the scalar field can be screened in the high density environment and evade the tight constraints given by solar system experiments.
However, in the large scale, the matter density is low, the scalar field can have significant effects to accelerate the expansion of the universe.
The mass of the compact object $m_a(\phi)$ can also be expanded as
\bq
m_a(\phi) = m_a\Bigg[1+s_a\left(\frac{\delta\phi}{\phi_{\rm VEV}}\right)+\mathcal{O}\left(\frac{\delta\phi}{\phi_{\rm VEV}}\right)^2\Bigg],
\eq
where $m_a=m_a(\phi_{\rm VEV})$. The sensitivity of the $a$-th object $s_a$ is defined as
\bq
s_a \equiv \frac{\partial(\ln m_a)}{\partial(\ln \phi)}\bigg|_{\phi_{\rm VEV}}.
\eq
In most cases, the deviations from GR are quantified by the sensitivity \citep{PhysRevD.93.124003,zhang2017gravitational,liu2018waveforms,PhysRevD.100.024038,Zhang2019Constraints}. In the SMG, it is proportional to the screened parameter
\bq
s_a=\frac{\phi_{\rm VEV}}{2 M_{\rm Pl}}\epsilon_a,
\eq
and the screened parameter of a uniform density object is given by
\bq
\epsilon_a=\frac{\phi_{\rm VEV}-\phi_a}{M_{\rm Pl}\Phi_a},
\eq
where $\Phi_a=Gm_a/R_a$ is the surface gravitational potential, and $\phi_a$ is the position of the minimum of the effective potential inside the object.
In this paper, we will focus on the screened parameter and forecast how tight constraints can be placed on it by the future space-borne GW detectors.

In the wave zone, the linear field equations are given by
\bq
\square\bar{h}_{\mu\nu} = -16\pi G\tau_{\mu\nu},
\eq
\bq
\left(\square-m^2_s\right)\delta\phi = -16\pi GS,
\eq
where $\bar{h}_{\mu\nu}=h_{\mu\nu}-\frac12\eta_{\mu\nu}h^{\lambda}_{\lambda}$, $\tau_{\mu\nu}$ is the total energy-momentum tensor and $S$ is the source term of the scalar field.
The solutions of these equations in the wave zone can be obtained by using the Green's function method, which are expressed in terms of the mass multipole moments and the scalar multipole moments. Based on the solutions, the GW waveforms in the SMG were calculated in the previous work \citep{liu2018waveforms}.

As shown in \citep{liu2018waveforms}, in addition to $+$ and $\times$ polarization modes in GR, the massive scalar field induces two polarizations, i.e. breathing polarization $h_b$ and longitudinal polarization $h_l$. The response of an interferometric detector is given by
\bq \label{response_f}
h(t) = F_+ h_+ + F_\times h_\times + F_b h_b + F_l h_l
\eq
where $F_{+, \times, b,\, l}$ denote the antenna pattern functions depending on the direction of GW sources ($\theta, \varphi$), detector configuration, polarization angle $\psi$, as well as the frequency of GWs for space-borne detectors, and $h_{+, \times, b,\, l}$ denote gravitational waveforms for the plus, cross, breathing and longitudinal polarization modes respectively.
Besides the same parameters in waveforms of GR, (which are total mass $m=m_1+m_2$, symmetric mass ratio $\eta=m_1m_2/(m_1+m_2)^2$, chirp mass $M_c=\eta^{3/5}m$, distance $D$, inclination angle $\iota$ between the line of sight and the binary orbital angular momentum, the time of coalescence $t_c$ and the orbital phase of coalescence $\Psi_c$), there are five extra parameters peculiar to SMG. They are the effective mass of the scalar field $m_s$, the expansion coefficients of the coupling function $A_0$ and $A_1$, the screened parameters of binary $\epsilon_1, \epsilon_2$ (see references \citep{zhang2017gravitational, liu2018waveforms} for more details). The Fourier transform can be obtained by using the stationary phase approximation.
The constraint $|A_0-1|$ is less than $10^{-10}$ according to the solar system experiments \citep{PhysRevD.93.124003}. Therefore, similar to \citep{liu2018waveforms}, we can safely adopt $A_0=1$ in our calculation. 
{As shown in the equation (74) of that paper \citep{liu2018waveforms}, the difference between the parameters in the Einstein frame and those in the Jordan frame is only a factor $A_0$. The parameters of the waveform in the Einstein frame and in the Jordan frame are same when we adpot $A_0=1$.
}
In addition, since the Compton wavelength $m_s^{-1}$ is roughly cosmological scale ($m_s^{-1}\sim1{\rm Mpc}$), as in the reference \citep{zhang2017gravitational} we set $m_s=0$ in the waveforms, which makes the $``l"$ polarization vanish ($\tilde{h}_l(f)=0$).
The results can be rewrote as following. The harmonic one is given by
\begin{equation} \label{hb_1}
\begin{aligned}
F_b\tilde{h}_b^{(1)}(f) ={} &\left(\frac{5}{48}\right)^{\frac12} \pi^{-\frac12} \frac{(G M_c)^{5/6}}{D}  (2f)^{-\frac76}
\Bigg[
-\frac{5}{384} E \epsilon_d^2 (2\pi f G m)^{-1}
+E(2\pi f G m)^{-\frac13}
\Bigg]  \\
&\times \exp \left\{
i \left[ 2\pi f t_c - \frac{\pi}{4} + \Psi(f) \right]
\right\},
\end{aligned}
\end{equation}

\begin{equation}
\begin{aligned}
F_l\tilde{h}_l^{(1)}(f) = 0 ,
\end{aligned}
\end{equation}
where
\begin{equation} \label{def_E}
E = -F_b A_1 M_{\rm Pl} \epsilon_d \sin \iota \left(1 + \frac12 \epsilon_1 \epsilon_2 \right)^{1/3},
\end{equation}
with $\epsilon_d=\epsilon_1-\epsilon_2$.
$\Psi(f)$ takes the form
\begin{equation}
\Psi(f) = -\Psi_c + \frac{3}{ 256 (2\pi f G M_c)^{5/3}} \sum_{i=-2}^{7} \Psi_i (2\pi f G m)^{i/3},
\end{equation}
where $\Psi_c$ is the orbital phase of coalescence, and the coefficients $\Psi_i$ are presented in appendix \ref{app_pn_waveform}. $\Psi_i(i\ge0)$ is the coefficients in 3.5PN phase function of Fourier domain waveform, and the coefficient $\Psi_{-2}$ is concerned with the correction of dipole radiation. The harmonic two is given by
\begin{equation} \label{hb_2}
\begin{aligned}
F_b\tilde{h}_b^{(2)}(f) ={} &\left(\frac{5}{96}\right)^{\frac12} \pi^{-\frac23} \frac{(G M_c)^{\frac56}}{D} f^{-\frac76} T
\Bigg[
F_b S_{-1} (\pi f G m)^{-\frac23}
+F_b
\Bigg] \\
&\times \exp \left\{ i[2\pi f t_c-\frac{\pi}{4}+2\Psi(f/2)] \right\},
\end{aligned}
\end{equation}

\begin{equation}
F_l\tilde{h}_l^{(2)}(f) = 0,
\end{equation}

\begin{equation} \label{h_p_c}
\begin{aligned}
F_+\tilde{h}^{(2)}_+(f) + F_{\times}\tilde{h}^{(2)}_{\times}(f) ={} &\left(\frac{5}{96}\right)^{\frac12} \pi^{-\frac23} \frac{(G M_c)^{\frac56}}{D} f^{-\frac76}
\Bigg[Q + Q S_{-1} (\pi f G m)^{-\frac23}\Bigg] e^{-i\varphi_{(2,0)}} P_{(2,0)}  \\
&\times  \exp \left\{ i[2\pi f t_c - \frac{\pi}{4}+2\Psi(f/2)] \right\},
\end{aligned}
\end{equation}
where
\begin{equation} \label{def_Q}
Q = \left(1 + \frac12\epsilon_1\epsilon_2\right)^{2/3},
\end{equation}
\begin{equation} \label{def_S-1}
S_{-1}= -\frac{5}{384}\epsilon_d^2,
\end{equation}
\begin{equation} \label{def_T}
T = -A_1 M_{\rm Pl} \Gamma \left(1+\frac12\epsilon_1\epsilon_2\right)^{2/3} \sin^2\iota,
\end{equation}
with $\Gamma = (\epsilon_1 m_2 + \epsilon_2 m_1)/m$.
And in the expression of $\tilde{h}^{(2)}_{+, \times}$, we have adopted the similar conventions of the reference \citep{van2006phenomenology}, where $e^{-i\varphi_{(2,0)}}P_{(2,0)}=-(1+\cos^2\iota)F_+ - i(2\cos\iota)F_{\times}$.

We also would like to investigate whether the constraints can be improved if the higher order amplitude corrections of the PN gravitational waveform are taken into consideration.
In the stage of adiabatic inspiral, the analytic waveforms can be got by using PN approximation where the waveforms can be expanded in terms of the orbital velocity. Thanks to the great efforts over the past few decades, the PN waveforms have been calculated to very high orders which are sufficiently precise to extract small signals buried in the large noise by matched filtering in GW experiments. More details can be found in the review article \citep{blanchet2014gravitational}.
Since the matched filtering method used in the GW detections is more sensitive to the phase of templates than the amplitude, the restricted waveform is the most commonly used waveform model, in which only the dominant harmonic is taken into account, except the leading order all amplitude corrections are discarded, but all the available order of phase are included. However, some works \citep{PhysRevD.77.024030, arun2007higher, Trias_2008, van2006phenomenology} have shown that it can induce considerable consequences if including higher order amplitude corrections in the templates. Here we consider the full PN waveform in which amplitude terms are included up to 2.5PN order and phase terms are included up to 3.5PN order.
The full waveform are shown in appendix \ref{app_pn_waveform} where we adopt the similar conventions of reference \citep{van2006phenomenology}.

\section{Space-borne gravitational wave detectors} \label{sec_detectors}
In this work, we consider two proposed space-borne GW detectors, Laser Interferometer Space Antenna (LISA) and TianQin, to forecast the constraints on SMG.
 LISA is a mission led by European Space Agency which can detect GWs in milli-Hz (0.1mHz-1Hz) range \citep{danzmann1996lisa, danzmann2016laser}. LISA consists of three identical spacecraft which maintain an equilateral triangular configuration in an Earth-trailing heliocentric orbit between 50 and 65 million km from Earth. The distance between two spacecraft is 2.5 million km according to the new LISA design \citep{danzmann2016laser}. The line connecting the Sun and the center-of-mass of the detector keeps a $60^\circ$ angle with respect to the plane of the constellation. Besides the revolution around the Sun, the detector rotates clockwise (viewed from the Sun) around its center-of-mass with a period of one year. The pictures depicting this orbit configuration can be found in Fig. 4.8 of reference \citep{danzmann1996lisa} or in Fig. 4 of reference \citep{danzmann2016laser}.
TianQin has the similar equilateral triangular configuration and is sensitive to the same frequency range. Different from LISA, TianQin is in a geocentric orbit with a period of 3.65 days. The distance between each pair of spacecraft is about $1.7\times10^5$ km. The normal vector of the detector plane is fixed and points toward the reference source J0806.3+1527 which is a candidate ultracompact WD binary in the Galaxy (${\rm longitude} = 120.5^\circ$, ${\rm latitude} = -4.7^\circ$ in the ecliptic coordinate system) and is a strong periodic GWs source in milli-Hz range. Illustrations of TianQin's configuration and orbit can be found in Fig.1 of the reference \citep{luo2016tianqin} or in Fig. 1 and A1 of \citep{hu2018fundamentals}.
There are two aspects of the detectors that are relevant to our analysis: the noise spectrum and the antenna pattern functions, which will be introduced respectively in the following subsections.

\subsection{Antenna beam-pattern functions}
In the proper detector frame, the response of a ground-based laser interferometer to GWs can be calculated by using the equation of the geodesic deviation,  which is \citep{maggiore2008gravitational, poisson2014gravity}
\bq \label{eq_geodesic_dev}
\ddot{\xi}^i = \frac{1}{2} \ddot{h}_{ij} \xi^j.
\eq
However, it is not completely correct to straightforwardly extend the same treatment to the situation of space-borne GW detectors. Since in the deriving of the equation of the geodesic deviation (\ref{eq_geodesic_dev}), the approximation that the distance between two test masses is much smaller than the typical scale over which the gravitational field changes significantly has be adopted. This means that the equation (\ref{eq_geodesic_dev}) can be used to derive the response function when the detector arm length is shorter than the reduced wavelength of the GWs (see section 1.3, 9.1 of reference \citep{maggiore2008gravitational}, and references \citep{hu2018fundamentals, cornish2003lisa} for more details).
The corresponding frequency which is called transfer frequency is given by
\bq
f_* = \frac{c}{2 \pi L},
\eq
where $L$ is the arm length of the detector. This condition is satisfied for the ground-based GW detectors, since the sensitive bands of ground-based detectors are far below their transfer frequency. While, it is not always satisfied for space-borne detectors. For instance, this critical frequency is $0.019\rm{Hz}$ for LISA, and it is $0.28\rm{Hz}$ for TianQin, which are similar to the sensitive frequency bands of the detectors.
The transfer frequencies of LISA and TianQin are illustrated by vertical dotted blue and orange lines respectively in Fig. \ref{fig_n_curve}.
For the science objectives such as supermassive black hole binaries, the equation in (\ref{eq_geodesic_dev}) can be safely used \citep{klein2016science,feng2019preliminary}. But in the case of BH-NS binaries considered in this work, the rough estimation of last stable frequency is much higher than the upper limit of detector's sensitive band. It is not proper to use the response function derived by extending the approach which is used for ground-based interferometers.

The propagation of GWs in the time during which the photons travel from laser source to photodetector is neglected if one use the equation (\ref{eq_geodesic_dev}) to derive the response function. When frequencies higher than the transfer frequency, there may be GWs of a few wavelengths passing through the path of photons during the time between emission and reception of the photons, which makes the effect of GWs to cancel out itself and deteriorates the response of detectors to the GWs.
In order to get the exact response of detectors, the integration along the null geodesic of photons between two test masses should be calculated.
The response functions of LISA-like detectors for two GR polarizations are given by references \citep{cornish2001space,cornish2003lisa, rubbo2004forward}, and the same process can be extended to other polarizations \citep{PhysRevD.99.104027}.

We present the general form of antenna pattern functions here, the explicit expression and the detail of the process can be found in Appendix \ref{app_pattern_f}.
The response of LISA or TianQin to GWs have shown in the Eq. (\ref{response_f}), where the antenna beam-pattern functions $F_{+, \times, b,\, l}$ are given by
\bq \label{pattern_f}
F_A = \frac{1}{2}\epsilon_{ij}^A
\left[
\hat{l}^i_1 \hat{l}^j_1 \, {\mathcal T}(f, \hat{\boldsymbol l}_1 \cdot \hat{\boldsymbol \Omega})
-\hat{l}^i_2 \hat{l}^j_2 \, {\mathcal T}(f, \hat{\boldsymbol l}_2 \cdot \hat{\boldsymbol \Omega})
\right],
\eq
with $A$ denoting different polarizations ($A=+, \times, b,\, l$), $\epsilon_{ij}^A$ denoting polarization tensors, $\hat{\boldsymbol l}_1$ and $\hat{\boldsymbol l}_2$ denoting the unit vectors of two arms.
Comparing the antenna beam-pattern functions of ground-based detectors, the differences are quantified by the transfer functions  ${\mathcal T}(f, \hat{\boldsymbol l}_1 \cdot \hat{\boldsymbol \Omega})$ and ${\mathcal T}(f, \hat{\boldsymbol l}_2 \cdot \hat{\boldsymbol \Omega})$, which are given by
\bq \label{transfer_f}
\begin{aligned}
{\mathcal T}(f, \hat{\boldsymbol l}_1 \cdot \hat{\boldsymbol \Omega}) ={} \frac{1}{2} \Biggl\{
&{\rm sinc} \left[ \frac{f}{2f_*}(1-\hat{\boldsymbol l}_1 \cdot \hat{\boldsymbol \Omega})\right]
\exp \left[ -i\frac{f}{2f_*}(3+\hat{\boldsymbol l}_1 \cdot \hat{\boldsymbol \Omega})\right]  \\
&+{\rm sinc} \left[ \frac{f}{2f_*}(1+\hat{\boldsymbol l}_1 \cdot \hat{\boldsymbol \Omega}) \right]
\exp \left[ -i\frac{f}{2f_*}(1+\hat{\boldsymbol l}_1 \cdot \hat{\boldsymbol \Omega})\right]
\Biggr\},
\end{aligned}
\eq
where ${\rm sinc}(x)=\frac{\sin x}{x}$, $\hat{\boldsymbol \Omega}$ denotes the unit vector of the GW propagation direction. In low frequency limit $f\ll f_*$, the transfer functions ${\mathcal T}(f, \hat{\boldsymbol l}_1 \cdot \hat{\boldsymbol \Omega})$ and ${\mathcal T}(f, \hat{\boldsymbol l}_2 \cdot \hat{\boldsymbol \Omega})$ approach to 1, which returns to the case of ground-based detectors.

For the equilateral triangular configuration, there are two independent output signals. The second output signal, following the previous work \citep{cutler1998angular}, is equivalent to the response of a two-arm detector rotated by $\pi/4$ with respect to the first one, in the assumptions that the noise is Gaussian, stationary and totally symmetric.
As shown in Appendix \ref{app_pattern_f}, $(\theta', \varphi', \psi')$ are employed to denote the GW source direction and the polarization angle in the detector coordinate. In terms of $(\theta', \varphi', \psi')$, the two output signals can be expressed as
\bq
\begin{aligned}
h^{I}(t) &= \sum_A F_A(\theta', \varphi', \psi') h_A , \\
h^{II}(t) &= \sum_A F_A(\theta', \varphi'-\pi/4, \psi') h_A ,
\end{aligned}
\eq
where $A=(+, \times, b,\, l)$.

\subsection{Noise spectra of GW detectors}
The noise of a GW detector can be characterized by the one-side noise power spectral density (PSD) $S_n(h)$. We employ the noise curve of LISA in the reference \citep{belgacem2019testing}, i.e.
\bq
S_n(f) = \frac{4S_{\rm acc}(f) + S_{\rm other}}{L_{\rm LISA}^2} \Biggr[ 1+\left(\frac{f}{1.29f_*}\right)^2 \Biggl] + S_{\rm conf}(f)
\eq
Here $f_*=0.019\rm{Hz}$ is the transfer frequency of LISA. $L_{\rm LISA}$ is the arm length which is 2.5 million km according to the new LISA design \citep{danzmann2016laser}.
The acceleration noise is given by
\bq
S_{\rm acc}(f) = \frac{9 \times 10^{-30} {\rm m}^2 {\rm Hz}^3}{(2 \pi f)^4}
\Biggl[1 +
\Biggr(\frac{6 \times 10^{-4}{\rm Hz}}{f}\Biggl)^2
\Biggr(1+\Bigr(\frac{2.22 \times 10^{-5}{\rm Hz}}{f}\Bigl)^8\Biggl) \Biggr].
\eq
The other noise is given by
\bq
S_{\rm other} = 8.899 \times 10^{-23} {\rm m}^2 {\rm Hz}^{-1}.
\eq
In addition to the noise from instruments, the numerous compact WD binaries in the Galaxy can emit GWs of a few mHz and produce the confusion noise. The confusion noise from unresolved binaries is approximated by
\bq
S_{\rm conf}(f) = \frac{A}{2} e^{-s_1 f^{\alpha}} f^{-7/3} \biggl\{ 1-{\rm tanh}\biggl[s_2(f-\kappa)\biggr] \biggr\},
\eq
with $A = (3/20)3.2665 \times 10^{-44} {\rm Hz}^{4/3}$,
$s_1 = 3014.3{\rm Hz}^{-\alpha}$,
$\alpha = 1.183$,
$s_2 = 2957.7 {\rm Hz}^{-1}$,
and $\kappa = 2.0928 \times 10^{-3} {\rm Hz}$.

\begin{figure}[h]
\centering
  \includegraphics{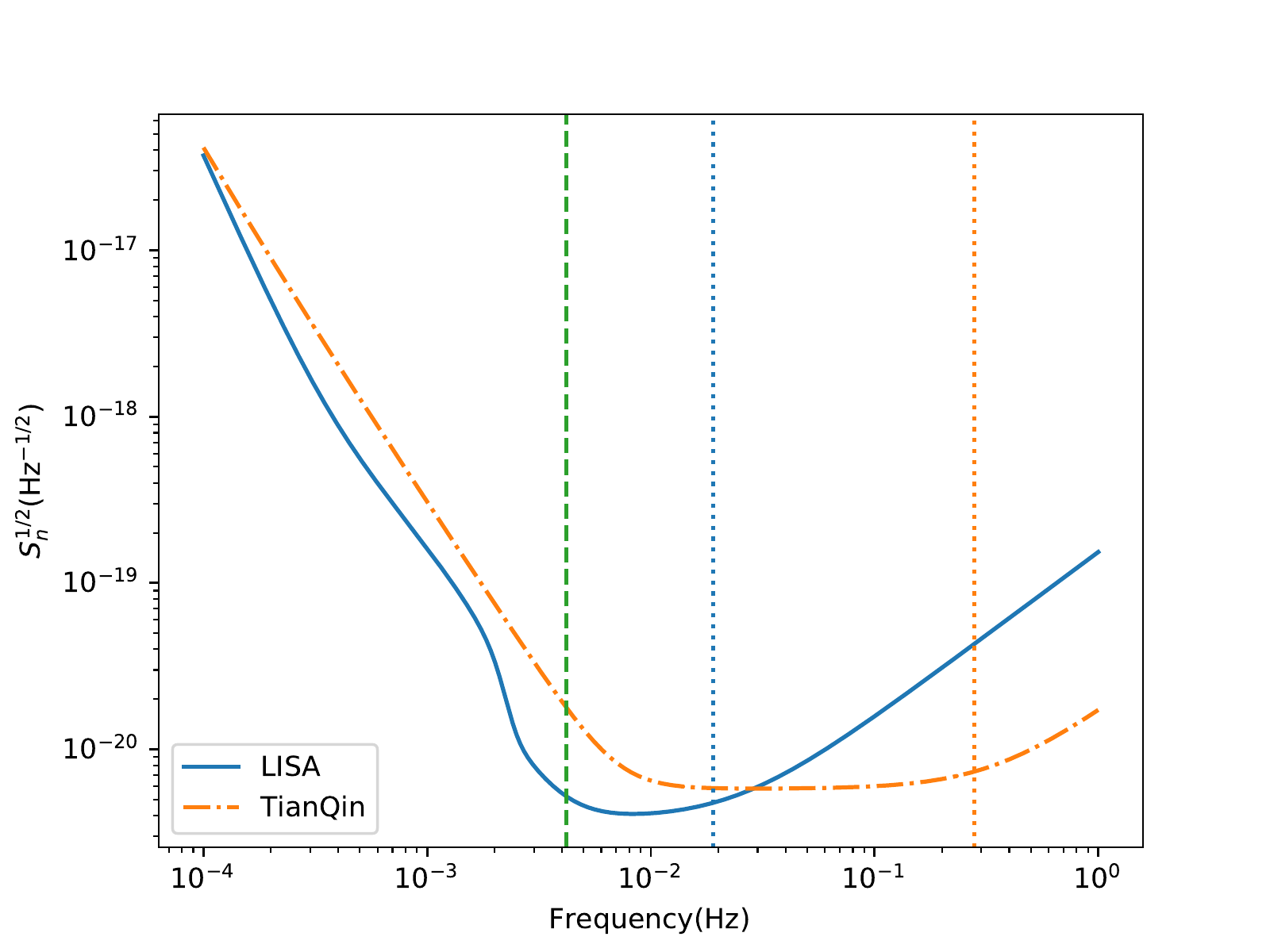}
  \caption{Sensitive curves of LISA and TianQin. The vertical dotted blue and orange lines denote the transfer frequency of LISA and TianQin respectively. The vertical dashed green line denotes the last stable orbital frequency of the BH-WD binaries used in this paper which is about $0.0042~\rm{Hz}$. For the BH-NS binaries, since the last stable orbital frequency is higher than the upper limit of the sensitive bands of both detectors, it is not shown in this figure.}
\label{fig_n_curve}
\end{figure}

For TianQin, we employ the noise curve provided in the reference \citep{luo2016tianqin,hu2018fundamentals,feng2019preliminary},
\bq
S_n(f) = \Biggl[ \frac{S_x}{L_{\rm TianQin}^2} + \frac{4S_a}{(2 \pi f)^4 L_{\rm TianQin}^2} \bigl(1+\frac{10^{-4}{\rm Hz}}{f} \bigr) \Biggr] \times \Biggl[ 1+(\frac{f}{1.29f_*})^2 \Biggr],
\eq
where $f_*=0.28\rm{Hz}$ is the transfer frequency of TianQin, $L_{\rm TianQin}=1.73 \times 10^8 {\rm m}$ is the arm length, $S_x = 10^{-24} {\rm m}^2/{\rm Hz}$ and $S_a = 10^{-30} {\rm m}^2{\rm s}^{-4}/{\rm Hz}$ are the position noise and acceleration noise respectively.
The sensitive curves $\sqrt{S_n(f)}$ of LISA and TianQin are presented in Fig. \ref{fig_n_curve}.

\section{Constraining the screened modified gravity} \label{sec_constr}

\subsection{Fisher information matrix} \label{subsec_fisher}
The Fisher matrix approach is widely used to estimate the precision of future experiments. Comparing with the techniques like Monte Carlo analysis, Fisher matrix is a simpler way to efficiently estimate errors of parameters in GW detection with sufficient accuracy in the high signal-to-noise ratio (SNR) cases \citep{finn1992detection,finn1993observing,cutler1994gravitational}.
The elements of Fisher matrix are given by
\bq \label{eq_Fisher_M}
\Lambda_{ij} = \left\langle \frac{\partial \tilde{h}(f)}{\partial p_i}, \frac{\partial \tilde{h}(f)}{\partial p_i} \right\rangle ,
\eq
where the $\tilde{h}(f)$ is Fourier transform of the output $h(t)$ of detectors, $p_i$ are the parameters to be estimated.
The angle brackets denote the detector-dependent inner product,
\bq \label{eq_inner_product}
\langle \tilde{a}(f), \tilde{b}(f) \rangle = 4\int_{f_1}^{f_2}\frac{\tilde{a}(f)\tilde{b}^*(f)+\tilde{a}^*(f)\tilde{b}(f)}{2}
\frac{df}{S_n(f)},
\eq
where $S_n(f)$ is the PSD of detector.
The upper limit of integral interval $f_2$ is determined by min($f_{\rm LSO}$, $f_{\rm up}$), {where $f_{\rm up}$ is the upper limit of the detector's sensitive band (1Hz)}, and $f_{\rm LSO}$ is the last stable orbital frequency of the binary which will be discussed in the section (\ref{constr}).
The lower limit of integral interval $f_1$ is given by max($f_{\rm low}$, $f_{\rm obs}$). Here the $f_{\rm low}$ is the lower limit of the detector's sensitive band ($10^{-4}$Hz). The $f_{\rm obs}$ corresponds to the orbital frequency at $T_{\rm obs}$ earlier from the time corresponding to $f_2$.
Approximately, $T_{\rm obs}$ can be regarded as the designed mission duration of the detector. For LISA $T_{\rm obs} = 4$ years \citep{danzmann2016laser}, and for TianQin $T_{\rm obs} = 5$ years \citep{luo2016tianqin}.
Using the formula of orbital decay to leading order (Eq. (\ref{orbital_dacy}) with $\epsilon_d=0$), the relation between orbital frequency at the beginning and ending of any time interval can be given by
\bq \label{f_obs}
f_{\rm obs}=f_2\left[1+\frac{256}{5}(GM_c)^{\frac53}T_{\rm obs}(2\pi f_2)^{\frac83}\right]^{-\frac38}.
\eq

The Fisher matrix for the combination of the two independent output signals is given by
\bq
\Lambda_{ij} = \Lambda^{I}_{ij}+\Lambda^{II}_{ij},
\eq
where $\Lambda^{I}_{ij} = \langle \frac{\partial \tilde{h}^{I}}{\partial p_i}, \frac{\partial \tilde{h}^{I}}{\partial p_i} \rangle$, $\Lambda^{II}_{ij} = \langle \frac{\partial \tilde{h}^{II}}{\partial p_i}, \frac{\partial \tilde{h}^{II}}{\partial p_i} \rangle$.
Using the definition of inner product, the combined SNR of two independent signals is
\bq \label{SNR}
\rho^2 = (\rho^{I})^2+(\rho^{II})^2,
\eq
with $(\rho^{I})^2 = \langle \tilde{h}^{I}, \tilde{h}^{I} \rangle$ and $(\rho^{II})^2 = \langle \tilde{h}^{II}, \tilde{h}^{II} \rangle$.
The covariance matrix $\Sigma$ can be derived by taking the inverse of Fisher matrix $\Lambda$, i.e.
$\Sigma = \Lambda^{-1}$, and
the estimation of the RMS error of a parameter $p_i$ is given by,
$\Delta p_i = \sqrt{\Sigma_{ii}}$.
The correlation coefficients between parameters $p_i$ and $p_j$ are given by
\bq
c_{ij} =\frac{\Sigma_{ij}}{(\Sigma_{ii}\Sigma_{jj})^{1/2}}.
\eq

\subsection{Constraining screened modified gravity} \label{constr}
As shown in references \citep{zhang2017gravitational,liu2018waveforms}, the orbital decay of compact binary, due to the gravitational radiation, is given by
\bq \label{orbital_dacy}
\dot{\omega}(t)=\frac{96}{5}(G M_c)^{\frac{5}{3}}\omega^{\frac{11}{3}}
\left[1+\frac{5}{192}(Gm\omega)^{-\frac23}\epsilon_d^2\right],
\eq
where $\omega$ is the orbital angular frequency and $\epsilon_d=\epsilon_1-\epsilon_2$ is the difference between the screened parameters of two objects.
We find that the asymmetric binary systems can induce more phase corrections, which induces the foremost difference between SMG and GR. Therefore, in this work, we consider the asymmetric systems, BH-NS binaries and BH-WD binaries, as the GW sources.

Since the stationary phase approximation, which is used to get the Fourier transform of a detector's response, is maintained only in the stage where the change of orbital frequency is negligible in the period of a single circle, and the PN waveforms are not accurate enough in the late stage, a specific frequency must be chosen where the waveforms are truncated.
As the rough estimations, we employ the Roche radius of rigid spherical body as last stable distance between the two objects \citep{PhysRevD.65.042002}, which are given by
\bq \label{roche_r}
d=R_{\rm comp} \left(2\frac{M_{\rm BH}}{M_{\rm comp}}\right)^\frac13.
\eq
Here, $R_{\rm comp}$ denotes the radius of the companion, $M_{\rm BH}$ and $M_{\rm comp}$ denote the mass of the black hole and the companion respectively.
The corresponding orbital frequency is given by
\bq \label{last_f}
f=\frac{1}{2\pi}\sqrt{\frac{G(M_{\rm BH}+M_{\rm comp})}{d^3}}.
\eq
For the NS, the typical values of mass and radius are $M_{\rm NS}=1.4 {\rm M}_\odot$ and $R_{\rm NS}=10 {\rm km}$. For the WD, we employ the parameters of target in PSR J1738+0333, which are given by $M_{\rm WD}=0.181 {\rm M}_\odot$, $R_{\rm NS}=0.037 {\rm R}_\odot$ \citep{antoniadis2012relativistic}.
For the massive BH in the binary systems, we consider two cases in this paper, $M_{\rm BH}=1000 {\rm M}_\odot$ and $M_{\rm BH}=10000 {\rm M}_\odot$.
The last stable orbital frequencies of BH-WD binaries are approximate 0.0042 Hz for the both binaries with different BH mass, which are illustrated by the vertical dashed green line in Fig. \ref{fig_n_curve}. Since the $f_{\rm LSO}$ of BH-NS binaries exceeds the upper limit of detectors' sensitive band, it is not shown in Fig. \ref{fig_n_curve}.
Similar to the previous works \citep{Will2004testing},
the locations of the GW sources are set to Virgo cluster, where the distance is $D=16.5 {\rm Mpc}$ and the celestial position in the ecliptic coordinate system is longitude $181.04^\circ$ latitude $14.33^\circ$.

Since the screened parameter of BH is zero \citep{ liu2018waveforms}, the terms related to $\epsilon_1\epsilon_2$ vanish and $|\epsilon_d|$ becomes the screened parameter of the companion $\epsilon_{\rm NS}$ or $\epsilon_{\rm WD}$. Since the GW waveforms contain the term $\epsilon^2_{\rm NS, WD}$, rather than $\epsilon_{\rm NS, WD}$. For this reason, similar to \citep{liu2018waveforms, Zhang2019Constraints}, we constrain the parameter $\epsilon^2_{\rm NS, WD}$, instead of $\epsilon_{\rm NS, WD}$ in our analysis. Besides, since $A_1$ and $\epsilon_{d}$ always appear together in the GW waveforms, to evade singular matrix in the practical computations of Fisher matrix, we consider $A_1\epsilon_{d}$ as a combination and constrain it, instead of the parameter $A_1$.

\begin{figure}[h]
\centering
  \includegraphics[scale=0.5]{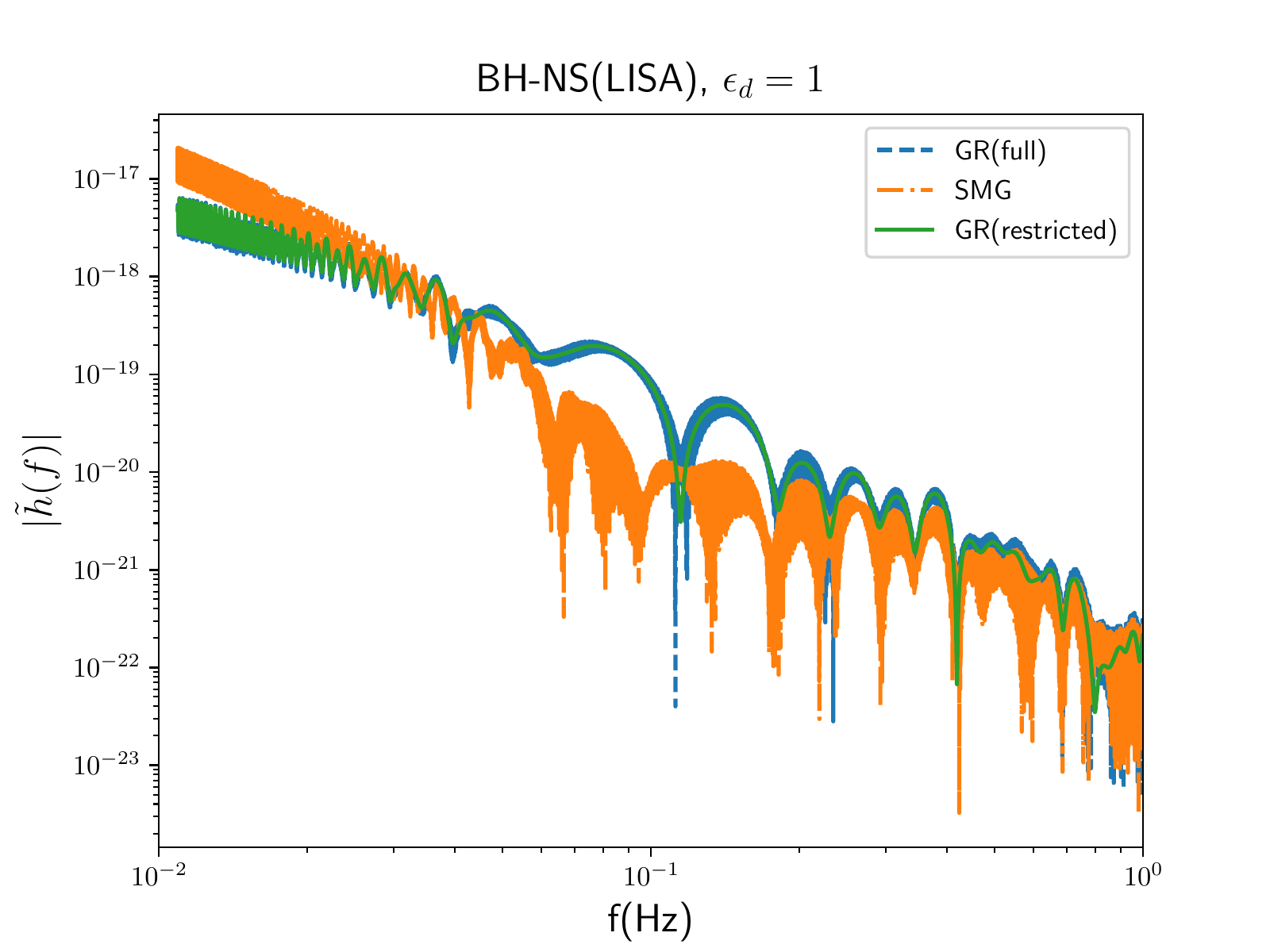}
  \includegraphics[scale=0.5]{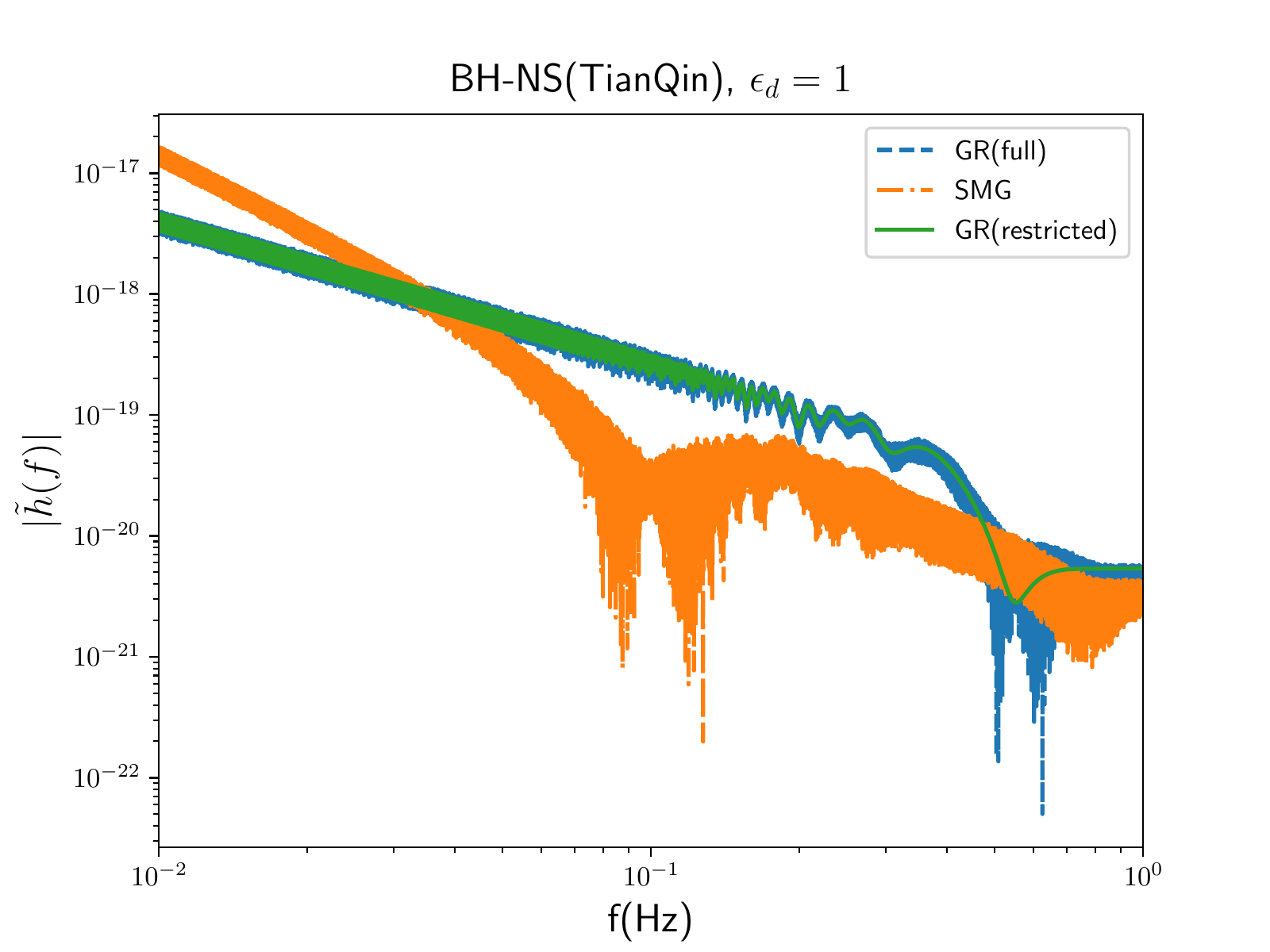}
  \includegraphics[scale=0.5]{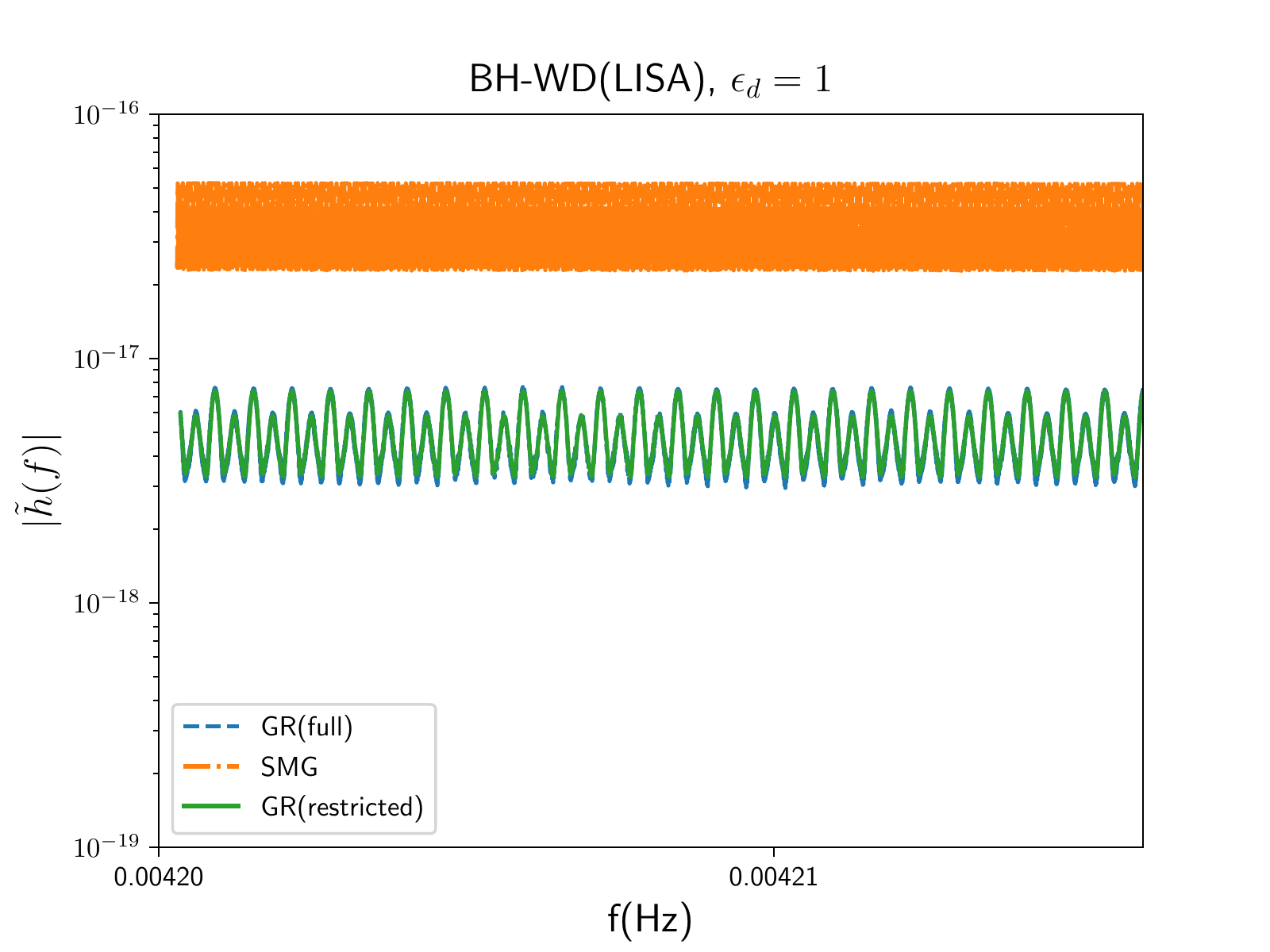}
  \includegraphics[scale=0.5]{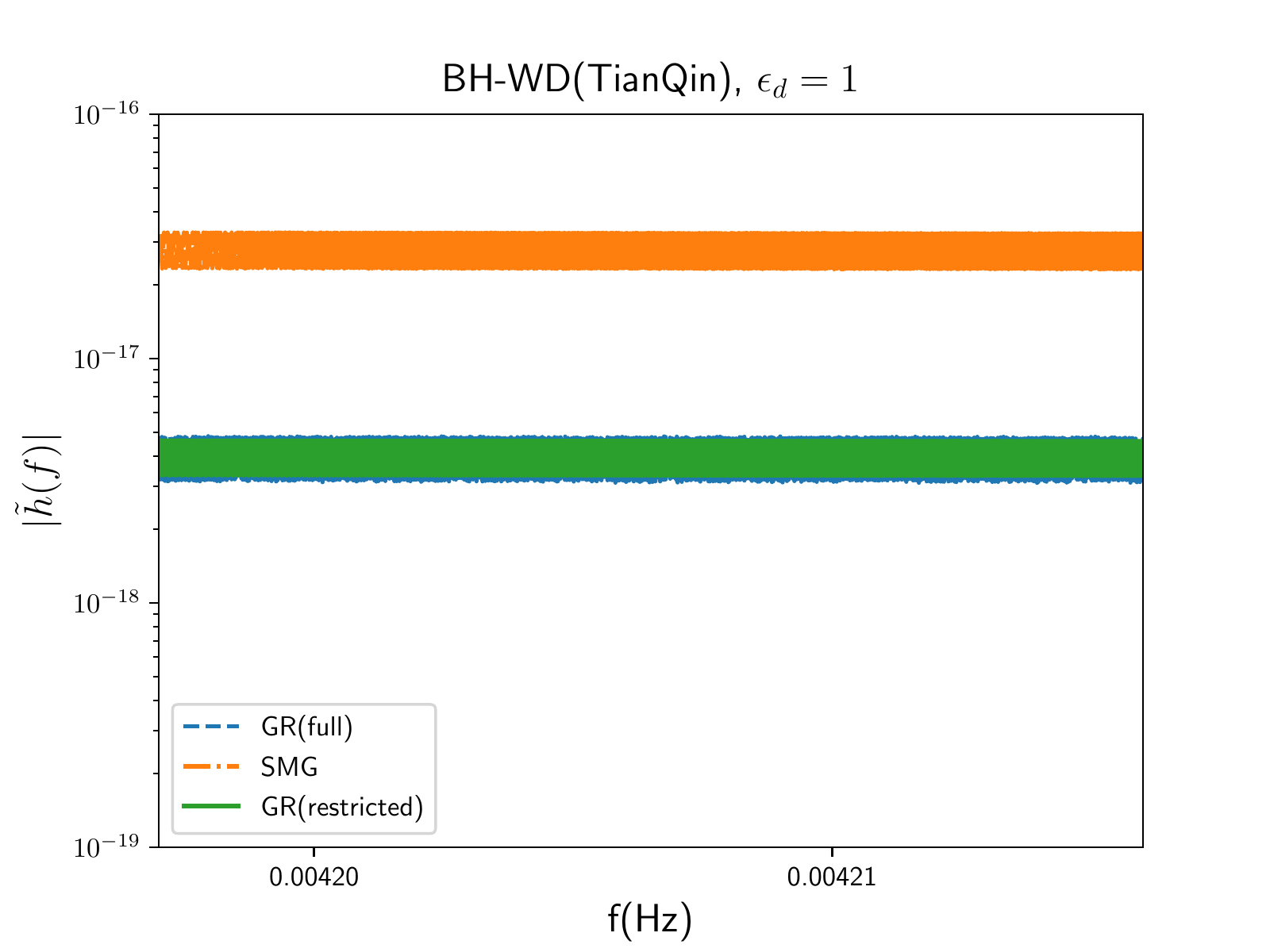}
  \caption{The detectors' responses $\tilde{h}(f)$ to the GWs, produced by BH-NS binaries and BH-WD binary, where we have set $m_{\rm BH}=1000M_{\odot}$ and $\iota=45^{\circ}$ in this figure. In each panel, the dashed blue line denotes the full waveforms in GR, the dash-dotted orange line denotes the full waveforms in SMG, and the solid green line denotes the restricted waveforms in GR. In order to show the deviation from GR, we consider the extreme case with $\epsilon_d=1$ in this picture. The frequency intervals are determined by the rules discussed in section \ref{subsec_fisher}. The waveforms are truncated by last stable frequency. For the BH-WD binary, this frequency is about 0.0042Hz. For the BH-NS, this frequency is higher than the upper limit of detectors' sensitive band. The lower band of frequency are determined by the designed mission duration of detectors. The frequency intervals of BH-WD binaries are very small where the orbital frequency of binaries has hardly any changes. Since the full waveforms in GR are close to the restricted waveforms, especially at low frequency, the blue lines denoting the full waveforms in GR are overlapped with the green lines denoting the restricted waveforms. The oscillations in low frequency are induced by the motion of space-borne detectors, which are absent for ground-based detectors. And the irregular fluctuations in high frequency are owing to the transfer functions in Eq. (\ref{transfer_f}).}
\label{waveform_BH-WD}
\end{figure}

In the Fisher matrix analysis, there are 11 free parameters in the full response functions, which are
\bq \label{paras}
\left(\iota, \ln M_c, \ln \eta, \ln r, t_c, \Psi_c, \theta, \varphi, \psi, A_1 M_{\rm Pl}\epsilon_{d}, \epsilon^2_{d} \right).
\eq
As mentioned above, the RMS errors of the parameters can be estimated by the Fisher matrix method for future GW experiments. The GW waveforms return to those of GR, when $A_1=0$ and $\epsilon_{\rm NS, WD}=0$. So, in this article, we set $A_1=\epsilon_{\rm NS, WD}=0$ in the fiducial waveforms, and their RMS errors can be considered as the upper limits of $A_1$ and $\epsilon_{\rm NS, WD}$ by the potential observations. The values of other parameters in fiducial waveforms are set as $\left(t_c=0, \Psi_c=0, \psi=0\right)$.
The fiducial waveforms are presented in Fig. \ref{waveform_BH-WD}, where we have set the inclination angle $\iota=45^{\circ}$ and the mass of BH is chosen to be $1000M_{\odot}$.

In our analysis, we consider three different cases to investigate the capabilities of LISA and TianQin.
\begin{enumerate}
\item In the first case, we assume the GW detectors will constrain SMG only by observing the extra polarization modes of GWs. As mentioned in the previous discussion, it includes only the breathing polarization mode $h_b$, since we have set the mass of scalar field $m_s\rightarrow 0$ in our calculation. The expansion coefficient of the coupling function $A_1$ is set to a specific value.
The $f(R)$ gravity can be cast into the form of a scalar-tensor theory, and scalar degree of freedom can be suppressed in high-density regions by the chameleon mechanism. The coupling function in $f(R)$ gravity is given by $A(\phi)=\frac{1}{\sqrt{f'(R)}}=\exp(\frac{\xi\phi}{M_{\rm Pl}})$ with $\xi=1/\sqrt6$ \citep{liu2018constraining}.
Here, we choose $A_1 M_{\rm Pl}=1/\sqrt{6}$ as a characteristic value.
Thus, the amplitude of $h_b$ is quantified by the screened parameter $\epsilon_{\rm NS, WD}$, which will be constrained by the potential observation. We consider the values of $\epsilon_{\rm NS, WD}$ which can make the SNR reach to 10 as the constraints of screened parameters of NSs or WDs.

\item In the second case, we constrain the SMG by analyzing the deviation of GW waveform in SMG from that in GR. The Fisher matrix technique will be employed for analysis. Since the GW detectors are sensitive to the GW phases, rather than the amplitudes, in this case, only the phase correction induced by dipole radiation is taken into consideration. We consider a waveform model including the restricted waveforms where all phase corrections are included while all amplitude corrections except the leading order are discarded, and the phase correction induced by dipole radiation. As shown in the reference \citep{liu2018waveforms}, the standard PPE framework can be applied to the waveforms in SMG when we consider the two tensor polarizations $h_+$ and $h_\times$. The general form of the detector's response function is given by
\bq
\tilde{h}(f)=\tilde{h}_{\rm GR}(f)[1+\alpha(\pi M_c f)^{\frac{a}{3}}]e^{i\beta(\pi M_c f)^{\frac{b}{3}}},
\eq
where $\alpha, a, \beta, b$ are the four PPE parameters and $\tilde{h}_{\rm GR}(f)$ denotes the response function in GR. In this case, we only consider the non-GR correction in phase. The PPE parameters are taken to $\alpha=a=0, \beta=-\frac{5}{14336}\epsilon^2_d\eta^{2/5}, b=-7$, and the restricted waveforms are employed in $\tilde{h}_{\rm GR}(f)$.
Since the terms related to $A_1$ only present in the amplitude corrections which are discarded in this case, there are 10 parameters remaining in the Fisher matrix.

\item In the third case, the high orders amplitude corrections of PN waveform are included and all available correction terms of GW waveforms are taken into account. The full response functions are presented in Appendix \ref{app_pn_waveform}.
In order to investigate the influence of higher order amplitude corrections to the constraints of SMG, we use the waveform model which include 3.5 PN phase corrections, 2.5 PN amplitude corrections and corrections concerning with the SMG both in amplitude and phase to be the input signals of the Fisher matrix. The 11 parameters in Eq. (\ref{paras}) all exist in the Fisher matrix, and
we can obtain the constraints of both $A_1 M_{\rm Pl} \epsilon_d$ and $\epsilon_{\rm NS,WD}$ from Fisher matrix analysis.

\end{enumerate}

\section{Result and discussion} \label{sec_result}

\subsection{Constraints on the screened parameters of neutron star and white dwarf} \label{constraints}
Using the process discussed in last section, we forecast the potential constraints on $\epsilon_{\rm NS}$ and $\epsilon_{\rm WD}$ by future space-borne GW detectors. Applying the analysis to the BH-WD systems, we find the constraint on $\epsilon_{\rm WD}$ cannot be derived, since the values of SNR for these signals are all less than 10.
We can give an example in which GW signals from a BH-WD binary with $m_{\rm BH}=1000M_\odot$ are observed by LISA for four years. The last orbital frequency can be estimated by Eq. (\ref{last_f}), which is $0.004216$. The designed mission duration of LISA is four years. The orbital frequency corresponding to four years before the last orbit is $0.004200$ which is given by Eq. (\ref{f_obs}).
We can observe from Fig. \ref{waveform_BH-WD} that the signals observed in the whole mission duration of detectors are nearly sinusoidal.
Therefore, the integral interval of the inner product (\ref{eq_inner_product}) is about $10^{-5}$ order-of-magnitude for this example. The order-of-magnitude of the response $\tilde{h}(f)$ and the noise $|S_n(f)|$ can be roughly read out from Figs. \ref{waveform_BH-WD} and \ref{fig_n_curve} respectively, which are $10^{-17}$ and $10^{-20}$. The SNR of this signal detected by LISA can be roughly estimated as $\sqrt{10}$. In fact, the SNR of the other cases are also less than $10$. Therefore, we conclude that the GW signals of BH-WD considered in this paper cannot be detected by space-borne LISA, Taiji or TianQin missions, and the constraint on $\epsilon_{\rm WD}$ is not available.

\begin{table}[h]
\centering
	\begin{tabular}{c|c|cc|cc|c|cc|cc}
\hline
	& \multicolumn{5}{c|}{1000 $M_\odot$} & \multicolumn{5}{c}{10000 $M_\odot$} \\
	\hline
	$\iota(\rm deg)$	 & case 1 & case 2 & SNR & case 3	& SNR & case 1 & case 2	& SNR & case 3 & SNR \\
	\hline
	0.1 & 3.14 & $3.8\times10^{-5}$ &110& $3.3\times10^{-5}$ &110& 4.21 & $6.2\times10^{-5}$ &310& $5.4\times10^{-5}$ &300\\
	30	 & 0.73 & $3.7\times10^{-5}$ &99& $3.5\times10^{-5}$ &99& 0.16 & $6.6\times10^{-5}$ &270& $5.5\times10^{-5}$	&270\\
	45	 & 0.17 & $4.1\times10^{-5}$ &83& $3.6\times10^{-5}$ &83& 0.11 & $7.1\times10^{-5}$ &230& $5.8\times10^{-5}$ &220\\
	60	 & 0.14 & $4.6\times10^{-5}$ &65& $4.0\times10^{-5}$ &66& 0.09 & $8.0\times10^{-5}$ &180& $6.2\times10^{-5}$ &180\\
	90	 & 0.12 & $5.7\times10^{-5}$ &43& $4.7\times10^{-5}$ &43& 0.08 & $9.4\times10^{-5}$ &120& $7.3\times10^{-5}$ &120\\
	\hline
	\end{tabular}
\caption{Constraints on $\epsilon_{\rm NS}$ given by TianQin, where we consider two cases of black hole mass with various inclination angles. }
\label{res_TianQin}
\end{table}

\begin{table}[h]
\centering
	\begin{tabular}{c|c|cc|cc|c|cc|cc}
\hline
	& \multicolumn{5}{c|}{1000 $M_\odot$} & \multicolumn{5}{c}{10000 $M_\odot$} \\
	\hline
	$\iota(\rm deg)$	 & case 1 & case 2 & SNR & case 3	& SNR & case 1 & case 2	& SNR & case 3 & SNR \\
	\hline
	0.1 & 3.40 & $4.8\times10^{-5}$ &140& $4.2\times10^{-5}$ &140& 3.90 & $5.0\times10^{-5}$ &500& $4.1\times10^{-5}$	&480\\
	30	 & 0.77 & $5.2\times10^{-5}$ &110& $4.4\times10^{-5}$ &110& 0.13 & $5.4\times10^{-5}$ &430& $4.4\times10^{-5}$ &420\\
	45	 & 0.73 & $5.7\times10^{-5}$ &92& $4.6\times10^{-5}$ &95& 0.09 & $5.8\times10^{-5}$ &360& $4.6\times10^{-5}$ &350\\
	60	 &	 0.71 & $6.6\times10^{-5}$ &71& $5.0\times10^{-5}$ &72& 0.07 & $6.6\times10^{-5}$ &280& $5.0\times10^{-5}$ &270\\
	90	 &	 0.69 & $8.4\times10^{-5}$ &40& $6.1\times10^{-5}$ &41& 0.06 & $8.5\times10^{-5}$ &160& $6.2\times10^{-5}$ &150\\
	\hline
	\end{tabular}
\caption{Constraints on $\epsilon_{\rm NS}$ given by LISA, where we consider two cases of black hole mass with various inclination angles.}
\label{res_LISA}
\end{table}

\begin{table}[h]
\centering
	\begin{tabular}{c|c|c}
\hline
	$\iota(\rm deg)$	 & 1000$M_\odot$ & 10000$M_\odot$ \\
	\hline
	0.1 & 2.6                & 1.8                 \\
	30  & $9.0\times10^{-3}$ & $6.2\times10^{-3}$  \\
	45  & $6.3\times10^{-3}$ & $4.4\times10^{-3}$  \\
	60  & $5.1\times10^{-3}$ & $3.5\times10^{-3}$  \\
	90  & $4.4\times10^{-3}$ & $3.0\times10^{-3}$  \\
	\hline
	\end{tabular}
\caption{Constraints on $A_1 M_{\rm Pl} \epsilon_{\rm NS}$ given by TianQin.}
\label{res_TianQin_A1}
\end{table}

\begin{table}[h]
\centering
	\begin{tabular}{c|c|c}
\hline
	$\iota(\rm deg)$	 & 1000$M_\odot$ & 10000$M_\odot$ \\
	\hline
	0.1 & 3.5                & 1.5                 \\
	30  & $1.2\times10^{-2}$ & $5.4\times10^{-3}$  \\
	45  & $8.5\times10^{-3}$ & $3.8\times10^{-3}$  \\
	60  & $6.9\times10^{-3}$ & $3.1\times10^{-3}$  \\
	90  & $6.0\times10^{-3}$ & $2.6\times10^{-3}$  \\
	\hline
	\end{tabular}
\caption{Constraints on $A_1 M_{\rm Pl} \epsilon_{\rm NS}$ given by LISA. }
\label{res_LISA_A1}
\end{table}

Let us turn to the cases with BH-NS binaries as the GW sources. We considered two kinds of binaries with different BH mass, i.e., $m_{\rm BH}=1000M_{\odot}$ and $m_{\rm BH}=10000M_{\odot}$, and for each case we consider the different inclination angles of the binary system. As shown in Eqs. (\ref{hb_1}) and (\ref{hb_2}), we find that the contributions of non-GR polarization induced by SMG to detector's response depend on $\iota$ by $\sin$ function, and
the polarization $h_b$ vanish when $\iota=0^{\circ}$. So, in order to avoid singularity, we choose $\iota=0.1^{\circ}$ instead of $\iota=0^{\circ}$ in the analysis.

The constraints of screened parameter for the three cases are present in Tables \ref{res_TianQin} and \ref{res_LISA} for TianQin and LISA respectively. From these results, we find the constraint of parameter $\epsilon_{\rm NS}$ is quite loose in case I, where only the extra breathing mode is used to constrain the SMG theory. Since the amplitude of this mode is much smaller than that of plus and cross modes, its contribution to the GW waveform modification is sub-dominant \citep{liu2018waveforms}. For this reason, although the production of extra polarization mode is a significant non-GR effect, it is hard to be detected in the actual observations. However, in case II and case III, the constraints $\epsilon_{\rm NS}\sim \mathcal{O}(10^{-5})$ are more than four orders tighter than that in case I. In addition, in comparison with case II and case III, we find that the constraints have only slightly improvement, if taking into account the contribution of high orders amplitude corrections of PN waveform. These results confirm the conclusions: for the test of SMG by space-borne detectors, the most important modification of GW waveforms are caused by the correction terms in GW phases, rather than by the extra polarization modes, or the correction terms in GW amplitudes.

For each case, we can compare the corresponding results of TianQin and LISA missions. For the case with same BH mass and inclination angle, we find that TianQin gives the better results for the cases of smaller BH mass (i.e. $m_{\rm BH}=1000M_{\odot}$), and LISA gives the better results for the cases of larger BH mass (i.e. $m_{\rm BH}=10000M_{\odot}$). Therefore, we conclude that, at least for constraining the SMG theory, TianQin is compatible for the smaller EMRIs, and LISA is compatible for the larger EMRIs. Meanwhile, by comparing the two cases of BH mass, we find that one can get tighter constraints from the binaries with lighter BH for TianQin, yet the difference between the two cases is not obvious for LISA. By observing the form of Fisher matrix (Eq. \ref{eq_Fisher_M} and \ref{eq_inner_product}), we can find that two kinds of information are inputted to the Fisher matrix. The one is the noise spectrum of a detector and the another is the partial differential of response $\tilde{h}(f)$ to different parameters which represents how the response $\tilde{h}(f)$ depends on a parameter. If the response $\tilde{h}(f)$ sensitively depends on a parameter, one can expect to the small RMS of this parameter or the tight constraint on this parameter.
For the parameter $\epsilon_{\rm NS}^2$, when derive the partial differential, we can find there is a factor $m_{\rm BH}^{-7/3}$ emerging where we have approximated $\eta \simeq \frac{m_{\rm NS}}{m_{\rm BH}}$ and $M \simeq m_{\rm BH}$ for EMSIs. Therefore, it is reasonable that the constraints on $\epsilon_{\rm NS}^2$ become loose when the BH mass increase.
Besides, the noise spectrum and antenna pattern function also influence the constraints on $\epsilon_{\rm NS}^2$. We attribute the regular pattern observed above to the different forms of two detectors' noise spectrum and antenna pattern function.

In case I with fixed BH mass, we can find that the constraints are looser for the smaller inclination angles, which is because the polarization $b$ depends on the inclination angle by sin function. For the case I, where only $b$ polarization is taken into account, the values of $\epsilon_{\rm NS}$ need to be higher when the inclination angle is small in order to make the SNR reach 10. However, in case II and case III with fixed BH mass, the smaller inclination angle follows the tighter parameter constraint, due to the fact that, relative to the edge-on sources, the face-on sources can be detected at the larger SNR.

In case III with the full GW waveform modifications, in addition to $\epsilon_{\rm NS}$, the model parameter $A_1$ can also be constrained. The results of $A_1$ are shown in Tables \ref{res_TianQin_A1} and \ref{res_LISA_A1}. We find that this parameter cannot be constrained well when the $\iota$ is too small. And the constraints are better for smaller inclination angle and heavier BH mass. LISA is more sensitive to the mass of BH. The constraints given by LISA are enhanced more when the mass of BH increases.

In summary, we find the best constraints expected to be reached by LISA mission are
$\epsilon_{\rm NS} \le 4.2\times10^{-5}$,
$A_1 M_{\rm Pl} \epsilon_d \le 6.0\times10^{-3}$
with $m_{\rm BH}=1000M_{\odot}$, and
$\epsilon_{\rm NS} \le 4.1\times10^{-5}$,
$A_1 M_{\rm Pl} \epsilon_d \le 2.6\times10^{-3}$
with $m_{\rm BH}=10000M_{\odot}$.
For TianQin, the forecasts are
$\epsilon_{\rm NS} \le 3.3\times10^{-5}$,
$A_1 M_{\rm Pl} \epsilon_d \le 4.4\times10^{-3}$
with $m_{\rm BH}=1000M_{\odot}$,  and
$\epsilon_{\rm NS} \le 5.4\times10^{-5}$,
$A_1 M_{\rm Pl} \epsilon_d \le 3.0\times10^{-3}$
with $m_{\rm BH}=10000M_{\odot}$, in the best case. Note that, in the previous work \citep{liu2018waveforms}, we have calculated the potential constraint of $\epsilon_{\rm NS}$ by the future ground-based Einstein telescope, and found that $\epsilon_{\rm NS}<6\times 10^{-4}(10^4/N_{\rm GW})^{1/4}$, where $N_{\rm GW}$ is the total number of GW events observed by Einstein telescope. Compared with this constraint, we find that constraints given by space-borne GW detectors are more than one order of magnitude tighter than those given by the third-generation ground-based GW detectors.

\subsection{Comparing with other observational constraints} \label{comb_consts}
In this section, we would like to compare the above results, which are forecasts for the future space-borne GW detectors, with the constraints placed by the present experiments, including pulsar timing observations, lunar laser ranging (LLR) and Cassini experiment.
{we find the constraint given by GW observations is complementary with the constraint from Cassini experiment, but weaker than those from LLR and binary pulsars.
Due to the strong surface gravitational potentials of neutron stars, although the screened parameter of neutron star can be constrained quite well, the constraint of the scalar background $\phi_{\rm VEV}$ is worse than that given by LLR. As for pulsar timing experiments of binary pulsars, the constraints on $\phi_{\rm VEV}$ are actually from the constraints on the orbital period decay caused by energy loss through gravitational radiation. This is similar to GW observations which constrain $\phi_{\rm VEV}$ by using the GW waveform. As we can see from the waveform (Eq.\ref{h1}, \ref{h2} and Eq.\ref{phase}), the terms concerning with SMG include the terms relevant to $(2\pi f G m)^{-1}$, $(2\pi f G m)^{-\frac13}$, $(\pi f G m)^{-\frac23}$ in amplitude (Eq.\ref{h1}, \ref{h2}) and $(2\pi f G m/n)^{-\frac23}$ in phase (Eq.\ref{phase}), which means that the SMG effects are more obvious in lower frequency range. The most sensitive frequency of space-borne GW detectors is about $10^{-2}{\rm Hz}$, while the orbital period of binary pulsars is in the order of 0.1 day. It is reasonable that the constraints given by pulsar timing are better than the constraints given by GW observations.
This result implies that at least for the three SMG models considered in this work, the GW observations by space-borne detectors may be not a good tool for the task of testing SMG theories.
}

Pulsar binary systems provide very useful tools to test gravity theories. The first indirect evidence of the existence of GW was given by the measurement of binary pulsar orbital period decay \citep{taylor1982new}. By monitoring the orbital period change, the deviation from GR can be constrained.
During the Apollo program and the Lunokhod missions, laser reflectors were installed on the moon. The laser pulses emitted on the earth can be reflected by the reflectors. By measuring the round-trip time, the earth-moon distance can be measured in extremely accuracy. The constraints on the Nordtvedt parameter and time variation of gravitational constant can be given by LLR experiment \citep{hofmann2010lunar}.
In this paper, we adopt the constraints in our previous work \citep{PhysRevD.100.024038}, which gave the upper bound on the scalar background $\phi_{\rm VEV}$ (the vacuum expectation value (VEV) of the scalar field in SMG) as follows,
\bq \label{c_pulsar}
\left(\frac{\phi_{\rm VEV}}{M_{\rm Pl}}\right)_{\rm pulsar} \le 4.4\times10^{-8},
\eq
by pulsar observations of PSRs J1738+0333 and J0348+0432 at $95.4\%$ confidence level (CL),
and the constraints by LLR at $95.4\%$ CL,
\bq \label{c_LLR}
\left(\frac{\phi_{\rm VEV}}{M_{\rm Pl}}\right)_{\rm LLR} \le 7.8\times10^{-15}.
\eq
The Cassini satellite was in solar conjunction in 2002. The Shapiro time-delay measurements using the Cassini spacecraft yielded a very tight constraint on the PPN parameter $\gamma$ \citep{ bertotti2003test}
\bq \label{c_Cassini}
|\gamma_{\rm obs}-1| \le 2.3\times10^{-5}.
\eq
These constraints will be compared with the potential constraint from future GW observations in this subsection.

In SMG, the screened parameter of a NS or WD can be approximated by \citep{zhang2017gravitational}
\bq \label{scr_parameter}
\epsilon_{a} = \frac{\phi_{\rm VEV}}{M_{\rm Pl}\Phi_{a}},
\eq
where $a$ denotes NS or WD, $\Phi_{a}=Gm_{a}/R_{a}$ is the surface gravitational potential of the $a$ object, and $\phi_{\rm VEV}$ is the scalar background in SMG. The constraints on the screened parameter $\epsilon_{a}$ can be converted to the constraints on the scalar background $\phi_{\rm VEV}$, and vice versa.
Here, we consider the best constraint on screened parameter given by TianQin, which are $\epsilon_{\rm NS} \le 3.3\times10^{-5}$, and compare with other observational constraints on SMG. The corresponding constraint on $\phi_{\rm VEV}$ given by TianQin is,
\bq  \label{c_GWNS}
\left(\frac{\phi_{\rm VEV}}{M_{\rm Pl}}\right)_{\rm GW(NS)} \le 7.6\times10^{-6}.
\eq
Similarly, we also consider the best result of $A_1$, i.e.,
\bq \label{c_A1}
A_1 M_{\rm Pl} \epsilon_{\rm NS} \le 2.6 \times 10^{-3},
\eq
as a typical value of constraint on $A_1$ to compare with other constraints.
In the following, we will compare these constraints in three specific SMG models: chameleon, symmetron and dilation theories.

\subsubsection{Chameleon}
\begin{figure}
\centering
	\includegraphics{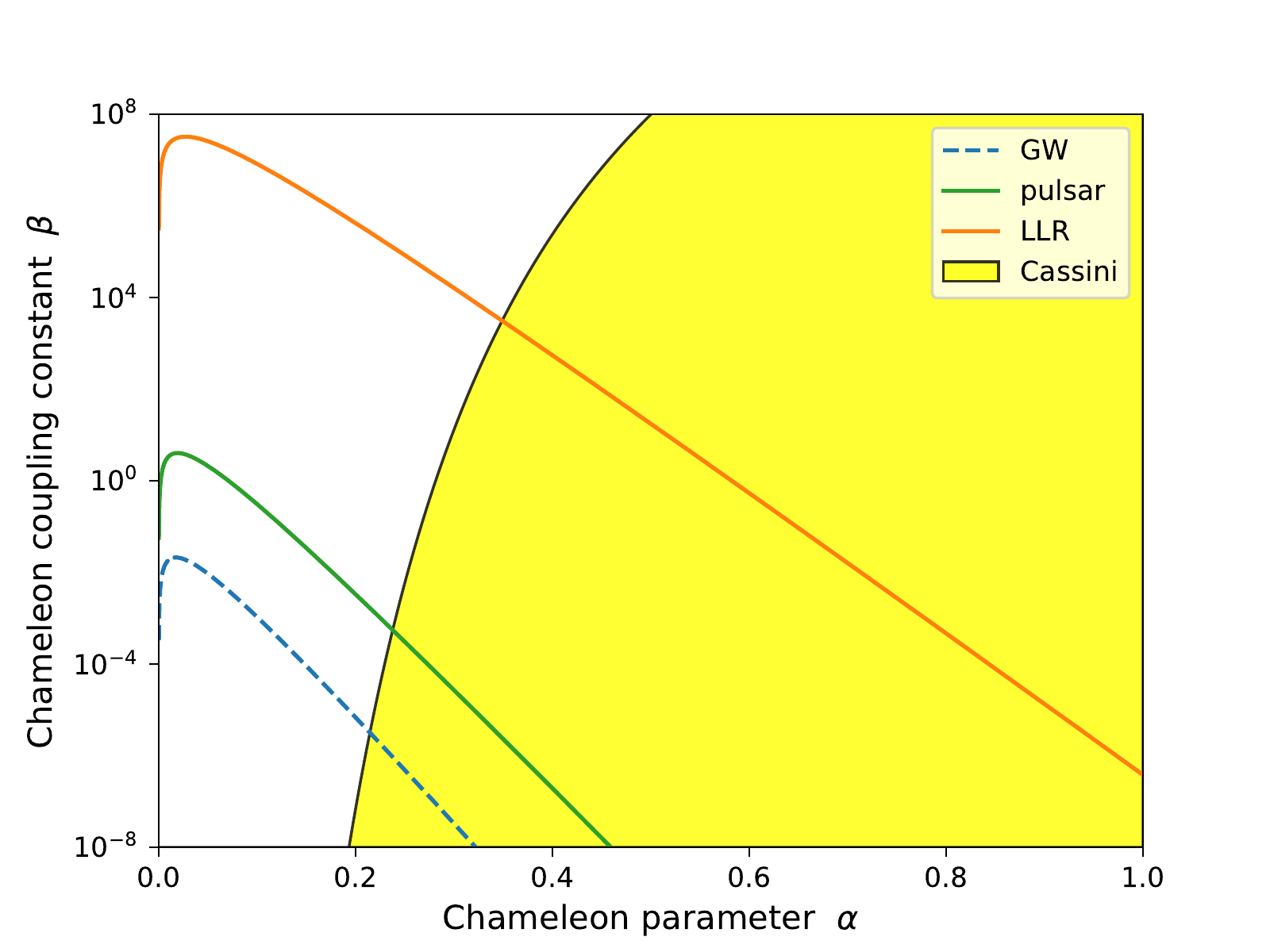}
\caption{The parameter space of exponential chameleon model. The dashed blue line denotes the forecast of constraints given by future space-borne GW detectors.
The solid green and orange lines denote the constraints given by the real experiments, pulsar observations and LLR, respectively. The allowed regions are the right areas of corresponding lines. The allowed region of constraint given by Cassini experiment is illustrated by the yellow shadow. }
\label{chameleon}
\end{figure}

The chameleon model was proposed by Khoury and Weltman \citep{khoury2004chameleonPRL, khoury2004chameleonPRD}, which introduced the screening mechanism by making the mass of scalar field depend on the environment density. The original chameleon model has been ruled out by the combined constraints from the Solar system and cosmology \citep{hees2012combined, PhysRevD.93.124003}. The idea of chameleon can be revived by introducing a potential and coupling function which have an exponential form \citep{brax2004detecting}
\bq
V(\phi)=\Lambda^4\exp(\frac{\Lambda^4\alpha}{\phi^\alpha}),~~~
A(\phi)=\exp(\frac{\beta\phi}{M_{\rm Pl}}).
\eq
Here $\alpha$, $\beta$ are positive dimensionless constants, $\Lambda$ denotes the energy scale of the theory which is required by the cosmological constraints to be close to the dark energy scale $2.24\times10^{-3}$eV \citep{PhysRevD.93.124003, hamilton2015atom}.
The scalar background $\phi_{\rm VEV}$ in chameleon model is given by \citep{zhang2017gravitational, PhysRevD.93.124003}
\bq \label{cha_vev}
\phi_{\rm VEV}=\left[\frac{\alpha M_{\rm Pl} \Lambda^{4+\alpha}}{\beta \rho_b} \right]^{\frac{1}{1+\alpha}},
\eq
where $\rho_b$ is the background matter density corresponding to the galactic matter density $\rho_{gal} \simeq 10^{-42}{\rm GeV}^4$ \citep{zhang2017gravitational}.
The PPN parameter $\gamma$ in chameleon model is given by \citep{PhysRevD.93.124003}
\bq \label{cha_gamma}
\gamma = 1-\frac{2\beta\phi_{\rm VEV}}{M_{\rm Pl}\Phi_{\rm Sun}}
\eq
where $\Phi_{\rm Sun}$ denotes the surface gravitational potential of the Sun.
The expansion coefficient of coupling function $A_1$ is given by \citep{PhysRevD.93.124003}, which is
${A_1}/{A_0} = {\beta}/{M_{\rm Pl}}$.
Recall that what we actually get by the process discussed above are the constraints on $A_1 M_{\rm Pl} \epsilon_{\rm NS}$ which takes the form of
\bq \label{cha_A1}
A_1 M_{\rm Pl} \epsilon_{\rm NS} = \beta \frac{\phi_{\rm VEV}}{M_{\rm Pl}\Phi_{\rm NS}},
\eq
where we have adopted $A_0=1$.
Using above formulae (Eqs. \ref{cha_vev}, \ref{cha_gamma}, \ref{cha_A1}) the constraints on the parameters of chameleon model can be obtained by the constraints on $\phi_{\rm VEV}$, $A_1$ and $\gamma$ (Eqs. \ref{c_pulsar}, \ref{c_LLR}, \ref{c_GWNS}, \ref{c_Cassini}, \ref{c_A1}).
We find that the express of $A_1$ in Eq. (\ref{cha_A1}) is similar to that of PPN parameter $\gamma$ in Eq. (\ref{cha_gamma}). The comparison can be glimpsed by comparing $\beta \phi_{\rm VEV}/M_{\rm Pl}$, which is
\bq
\left(\frac{\beta \phi_{\rm VEV}}{M_{\rm Pl}}\right)_{A_1} \le 5.4 \times 10^{-4},~~~
\left(\frac{\beta \phi_{\rm VEV}}{M_{\rm Pl}}\right)_{\rm Cassini} \le 2.4 \times 10^{-11}.
\eq
Since the constraint given by $A_1$ is much weaker than that derived from other observations, the allowed range will fill the full region of Fig. \ref{chameleon}. For this reason, the constraint corresponding to $A_1$ is not shown in Fig. \ref{chameleon}.

The other four constraints are illustrated in Fig. \ref{chameleon}, where the dashed line denotes the forecast for GW constraint, the solid lines denote the constraints of real experiments (pulsar and LLR), their allowed regions are the right areas of corresponding lines, and the region allowed by Cassini experiment is illustrated by the yellow shadow.
Although the GW observations can give the tight constraint on the screened parameter of NS, the constraint on the scalar background $\phi_{\rm VEV}$ cannot be improved simultaneously since the surface gravitational potential of NS is much larger than that of WD or solar system. We find that the most stringent bound on chameleon is still given by the combining constraint of LLR and Cassini  \citep{PhysRevD.100.024038} which gives $\alpha \ge 0.35$.

\subsubsection{Symmetron}
\begin{figure}
\centering
	\includegraphics{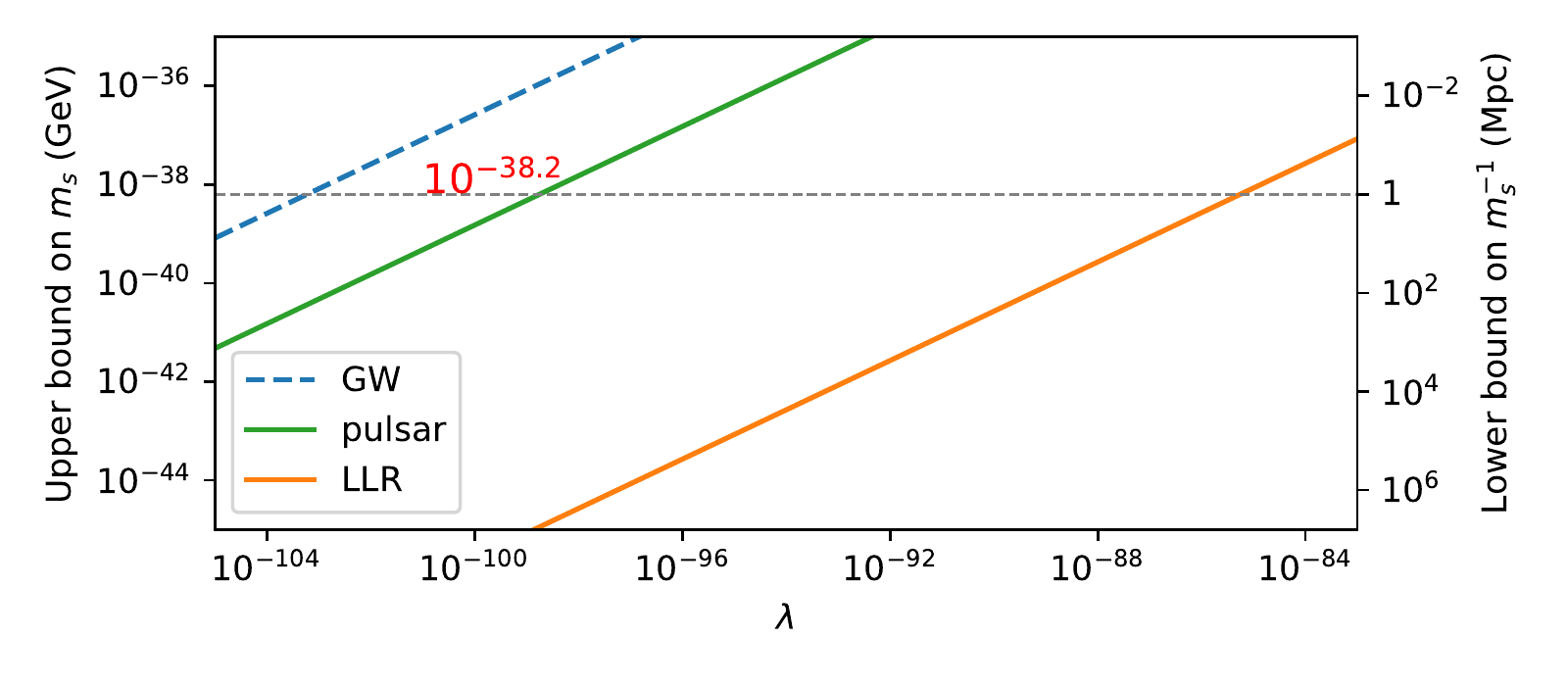}
	\includegraphics{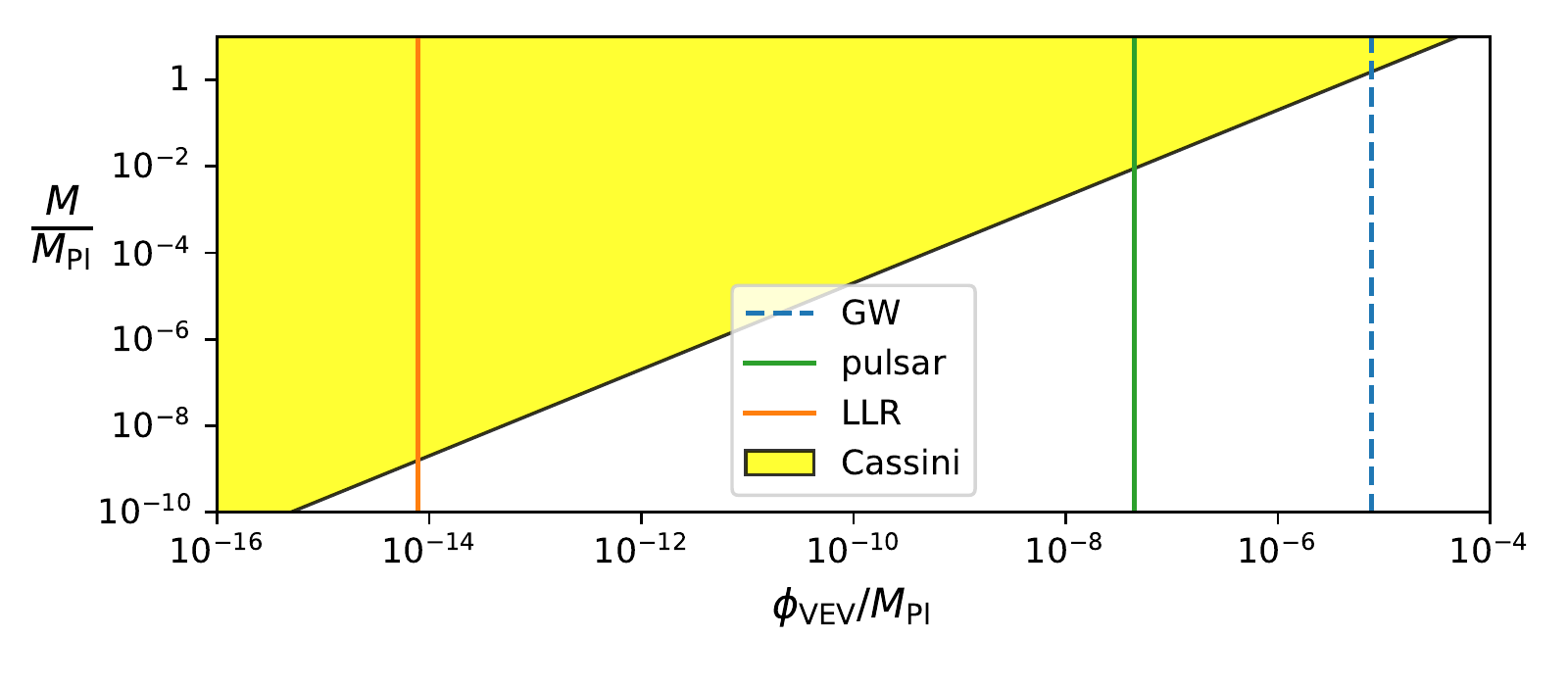}
\caption{Constraints on the symmetron model. The upper panel shows the upper bound on the scalar mass $m_s$ (or the lower bound on the $m_s^{-1}$) with the coupling constant $\lambda$. If $m_s^{-1}\le1$Mpc, the constraint on $\lambda$ given by LLR is $\lambda\ge10^{-85.3}$.
The bottom panel presents the constraints in parameter space ($\phi_{\rm VEV}$, $M$). The yellow area denotes the allowed region of constraint given by Cassini experiment. The dashed line denotes the forecast of the constraints given by future space-borne GW detectors, and the two solid lines denote the constraints given by pulsar and LLR. The corresponding allowed regions are the left areas of the lines.}
\label{symmetron}
\end{figure}

The symmetron models are characterized by a Mexican hat potential and a quadratic coupling function \citep{hinterbichler2010screening,hinterbichler2011symmetron,davis2012structure},
\bq
V(\phi)=V_0 - \frac{1}{2}\mu^2\phi^2 + \frac{\lambda}{4}\phi^4,
~~~
A(\phi)=1+\frac{\phi^2}{2M^2},
\eq
where $\mu$ and $M$ are mass scales, $\lambda$ is a positive dimensionless coupling constant, $V_0$ is the vacuum energy of the bare potential $V(\phi)$.
In the symmetron model, the VEV $\phi_{\rm VEV}$ is given by \citep{zhang2017gravitational, PhysRevD.93.124003},
\bq \label{symmetron_VEV}
\phi_{\rm VEV}=\frac{m_s}{\sqrt{2\lambda}},
\eq
which is proportional to the scalar mass. Similar to the chameleon model, the constraints on the scalar background $\phi_{\rm VEV}$ can be interpreted as the constraints on the parameters $m_s$ and $\lambda$ of symmetron model, which are shown in the upper panel of Fig. \ref{symmetron}.
The $m_s$ is the effective mass of the scalar field background. The scalar field background plays the role of dark energy which should have effects in large scales to accelerate the expansion of the universe. So the $m_s^{-1}$ is considered as roughly cosmological scales ($\sim 1$ Mpc) \citep{zhang2017gravitational}.

The PPN parameter $\gamma$ in symmetron model is \citep{zhang2017gravitational, PhysRevD.93.124003}
\bq
\gamma = 1-2\frac{\phi^2_{\rm VEV}}{M^2\Phi_{\rm Sun}}.
\eq
So, we can obtain the constraint on scalar background $\phi_{\rm VEV}$ with mass scale $M$ from the Cassini experiment, which is presented in the bottom panel of Fig. \ref{symmetron}.
The expansion coefficient of coupling function $A_1$ is given by \citep{PhysRevD.93.124003},
\bq
A_1 = \frac{\phi_{\rm VEV}}{M^2}.
\eq
And the term $A_1 M_{\rm Pl} \epsilon_{NS}$, that is treated as a parameter in the computations, takes the form of,
\bq
A_1 M_{\rm Pl} \epsilon_{\rm NS} = \frac{\phi^2_{\rm VEV}}{M^2\Phi_{\rm NS}}.
\eq
Thus, we can compare the constraints of $A_1$ with the constraints of Cassini experiment by comparing ${\phi^2_{\rm VEV}}/{M^2}$ directly,
\bq
\left( \frac{\phi^2_{\rm VEV}}{M^2} \right)_{A_1} \le 5.4 \times 10^{-4},
~~~
\left( \frac{\phi^2_{\rm VEV}}{M^2} \right)_{\rm Cassini} \le 2.4 \times 10^{-11}.
\eq
Since the surface gravitational potential of NS is much larger than that of the Sun, the constraint given by $A_1$ is much weaker than that given by Cassini experiment. The constraint given by $A_1$ is also not plotted in Fig. \ref{symmetron}.

The other four constraints are illustrated in Fig. \ref{symmetron}. The upper panel shows the upper bound on the mass of scalar field $m_s$ (or the lower bound on the $m_s^{-1}$) with the coupling constant $\lambda$. The dashed line denotes the forecast for GW constraints, the solid lines denote the constraints of real experiments (pulsar and LLR). The most stringent constraint is given by LLR. If $m_s^{-1}\le1$Mpc, the constraint on $\lambda$ given by LLR is $\lambda\ge10^{-85.3}$.
The bottom panel shows the constraints in parameter space ($\phi_{\rm VEV}$, $M$). The yellow area denotes the allowed region of constraint given by Cassini experiment. The dashed and solid vertical lines represent the constraints on the $\phi_{\rm VEV}$ given by the forecast of GWs (Eq. \ref{c_GWNS}) and the real experiments (Eq. \ref{c_LLR} and \ref{c_pulsar}) respectively. The corresponding allowed regions are the left areas of the lines. The most stringent constraint is still given by the combining constraint of LLR and Cassini.

\subsubsection{Dilaton}
In the dilation model, the potential and coupling function take the forms of \citep{damour1994stringNPB, damour1994stringGRG, brax2010dilaton},
\bq
V(\phi)=V_0 \exp(-\frac{\phi}{M_{\rm Pl}}),
~~~
A(\phi)=1 + \frac{(\phi-\phi_\star)^2}{2 M^2},
\eq
where $V_0$ is a constant which has the dimension of energy density, $M$ denotes the energy scale of the theory, and $\phi_\star$ represents the approximate value of scalar filed today.
The scalar background $\phi_{\rm VEV}$ and PPN parameter $\gamma$ are given by \citep{zhang2017gravitational, PhysRevD.93.124003}
\bq
\phi_{\rm VEV} = \phi_\star +\frac{M^2 \rho_{\Lambda_0}}{M_{\rm Pl} \rho_b},
~~~
\gamma = 1 - 2\frac{(\phi_{\rm VEV}-\phi_\star)^2}{M^2\Phi_{\rm Sun}}.
\eq
Here, $\rho_b$ is the background matter density, which is the galactic matter density $\rho_{gal} \simeq 10^{-42}{\rm GeV}^4$ in our calculation. $\rho_{\Lambda_0}$ denotes the density of dark energy, which is $\rho_{\Lambda_0} \simeq 2.51\times10^{-47}{\rm GeV}^4$ \citep{PhysRevD.93.124003}.
The screened parameter of a object takes the form of $\epsilon_a=\frac{\phi_{\rm VEV}-\phi_a}{M_{\rm Pl}\Phi_a}$, where $\phi_a=\phi_{\star}+\frac{M^2\rho_{\Lambda_0}}{M_{\rm Pl}\rho_a}$ is the minimum of the effective potential inside the object, and $\rho_a$ is the matter density inside the object.
Since the matter density in compact objects is much larger than that in cosmological background, we can drop the term $\frac{M^2\rho_{\Lambda_0}}{M_{\rm Pl}\rho_a}$ in the relation between the screened parameter $\epsilon_a$ and the parameter $M$ of dilaton model, which has the form of
\bq
\epsilon_a \Phi_a = \frac{M^2}{M_{\rm Pl}^2} \frac{\rho_{\Lambda_0}}{\rho_b}.
\eq
Another parameter which we can get the constraints on by GW observation takes the form of
\bq
A_1 M_{\rm Pl} \epsilon_{\rm NS} = (\frac{M}{M_{\rm Pl}})^2(\frac{\rho_{\Lambda_0}}{\rho_b})^2\frac{1}{\Phi_{\rm NS}}.
\eq
As in the previous models, the constraints on screened parameters $\epsilon_a$, $A_1$ and PPN parameter $\gamma$ can be switched to the constraints on the parameter $M$ of dilaton model.
The results are shown in Table \ref{dilaton}.
As mentioned before, although the future observations of GWs from BH-NS binaries can constrain the screened parameter of NS very stringently, the constraint on the mass scale $M$ by GW observations is not stringent, due to the large gravitational surface potential of NS. The tightest constraint on dilaton model is still given by LLR observation.
\begin{table}[h]
\centering
	\begin{tabular}{|c|c|c|c|c|c|}
	\hline
             & GW($\epsilon_{\rm NS}$) & GW($A_1$) & pulsar & LLR & Cassini \\
             	\hline
    $M/M_{\rm Pl}$ & $\le0.55$ & $\le920$ & $\le0.042$ & $\le1.8\times10^{-5}$ 	& $\le0.20$	\\
	\hline
	\end{tabular}
\caption{Constraints of the dilaton model derived from various observations. } \label{dilaton}
\end{table}

\section{Conclusion}

Gravitational wave provides an excellent opportunity to test GR, which is always considered as the most successful theory of gravity, in the strong-gravitational fields. In this issue, the calculation of GW waveforms in the alternative gravitational theory is important. The SMG is one of the simplest extension of GR in the scalar-tensor framework, which naturally explain the acceleration of cosmic expansion by introducing the scalar field. In addition, in this theory, the fifth force caused by the scalar field can be suppressed in the dense regions to satisfy various tests in solar system and laboratories. For these reasons, the SMG theory and its specific models, including chameleon, symmeton, dilaton, $f(R)$, etc., have been widely studied in the literature.
Based on the GW waveforms produced by the coalescence of compact binaries in general SMG derived in \citep{liu2018waveforms}, in this article we investigate the potential constraints on the general SMG theory by the future GW observations. In our calculation, we focus on the future space-borne missions, including LISA, TianQin and Taiji, and assume the EMRIs, including BH-NS and BH-WD in the Virgo cluster, as the GW targets.
By comparing three different cases, we find that the extra polarization modes, i.e. the breathing mode and the longitude mode, have little contribution in the constraining of model parameters. The modifications of GW waveforms in the plus and cross modes, in particular the correction terms in the GW phases, dominate the constraint of SMG parameters. If a GW signal produced by the coalescence of BH-NS system is detected by LISA, Taiji or TianQin, the screened parameter $\epsilon_{\rm NS}$ can be constrained at the level of $<\mathcal{O}(10^{-5})$. On the other hand, limited by the durations and the sensitive frequency bands of the GW detectors,
{we find that the GW signals produced by the coalescence of BH-WD system are difficult to be detected by LISA, Taiji or TianQin.}
For three specific SMG models: chameleon, symmeton, dilaton, we compare this potential constraint with the other existing constraints derived by Cassini experiment, LLR observations, and binary pulsars. We find that constraint from GW observation is complementary with that from Cassini experiment, but weaker than those from LLR observations, and binary pulsars.

~

~



\acknowledgments

This work is supported by NSFC Grants No. 11773028, No. 11633001, No. 11653002, No. 11421303, No. 11903030, No. 11903033, the Fundamental Research Funds for the Central Universities, the China Postdoctoral Science Foundation Grant No. 2019M652193, and the Strategic Priority Research Program of the Chinese Academy of Sciences Grant No. XDB23010200.



~~

~~

\appendix

\section{Post-Newtonian waveform} \label{app_pn_waveform}
{As shown in previous works \citep{PhysRevD.77.024030, arun2007higher, Trias_2008, van2006phenomenology}, for space-borne detectors, it can induce considerable consequences if higher order amplitude corrections are taken into consideration. In order to investigate whether including higher PN order amplitude corrections can affect the constraints on SMG or not, we adopt a waveform which combines the SMG corrections which have derived in our previous work \citep{liu2018waveforms} and the full GR waveform which is up to 3.5 PN in the phase and 2.5 PN in the amplitude.
In this Appendix, we present explicitly this waveform in which the corrections caused by SMG and the Doppler modulation peculiar to space-borne detectors are included.} We adopt the conventions similar to reference \citep{van2006phenomenology} where the full waveforms in GR have been presented.

In the PN approximation, the waveforms can be expressed as expansions in the typical internal speed of the source. The general forms of two GR polarizations are written as,
\begin{equation}
h_{+,\times}(t) = \frac{2m\eta}{D} x \left\{ H_{+,\times}^{(0)} + x^{1/2}H_{+,\times}^{(1/2)} + x^{1}H_{+,\times}^{(1)} + x^{3/2}H_{+,\times}^{(3/2)} + x^{2}H_{+,\times}^{(2)} + x^{5/2}H_{+,\times}^{(5/2)} \right\},
\end{equation}
where $x$ is the expansion parameter, which is defined as $x=[2\pi m F(t)]^{2/3}$ with $F(t)$ the orbital frequency. The expansion coefficients $H_{+,\times}^{(s)}$ consist of linear combinations of $\cos[n\Psi(t)]$ and $\sin[n\Psi(t)]$, where $\Psi(t)$ is the orbital phase and the number of harmonics $n=7$ for 2.5PN order in amplitude. The explicit expressions of $H_{+,\times}^{(s)}$ can be found in reference \citep{Arun_2004, Arun_2005}.
As mentioned in Eq. \ref{response_f}, the response function depends not only on the waveforms, but also on the antenna pattern functions. The analytic expressions for Fourier transform of a detector's response function can be got by using stationary phase approximation. The expressions are the sum of seven harmonics, i.e.
\bq \label{full_resf}
\tilde{h}(f) = \sum^7_{k=1} \tilde{h}^{(k)}(f).
\eq
Taking into account the corrections caused by SMG, the explicit expressions of $\tilde{h}^{(k)}(f)$ are presented as below,
\begin{equation} \label{h1}
\begin{aligned}
\tilde{h}^{(1)}(f) ={} &\left(\frac{5}{48}\right)^{\frac12}\pi^{-\frac23}\frac{(G M_c)^{5/6}}{D}(2f)^{-\frac76}
\Bigl\{
-\frac{5}{384} \pi^{\frac16} E \epsilon_d^2 (2\pi f G m)^{-1} \\
&+\pi^{\frac16} E(2\pi f G m)^{-\frac13} \\
&+e^{-i\varphi_{(1,1/2)}}P_{(1,1/2)}(2\pi f G m)^{\frac13} \\
&+ \left[ e^{-i\varphi_{(1,3/2)}}P_{(1,3/2)}+e^{-i\varphi_{(1,1/2)}}P_{(1,1/2)}S_1\right](2\pi f G m)  \\
&+ \left[ e^{-i\varphi_{(1,2)}}P_{(1,2)}+e^{-i\varphi_{(1,1/2)}}P_{(1,1/2)}S_{3/2} \right](2\pi f G m)^{\frac43}  \\
&+ \left[ e^{-i\varphi_{(1,5/2)}}P_{(1,5/2)}+e^{-i\varphi_{(1,3/2)}}P_{(1,3/2)}S_{1}+e^{-i\varphi_{(1,1/2)}}P_{(1,1/2)}S_{2} \right](2\pi f G m)^{\frac53}
\Bigr\}  \\
&\times \Theta(f_{\rm LSO}-f) \exp \left\{
i \left[ 2\pi ft_c-\frac{\pi}{4}+\Psi(f) \right]
\right\},
\end{aligned}
\end{equation}

\begin{equation} \label{h2}
\begin{aligned}
\tilde{h}^{(2)}(f) ={} &2^{-\frac12}\left(\frac{5}{48}\right)^{\frac12}\pi^{-\frac23}\frac{(G M_c)^{5/6}}{D}(f)^{-\frac76}
\Bigl\{
\left[ T F_b S_{-1} + Q S_{-1} e^{-i\varphi_{(2,0)}} P_{(2,0)} \right] (\pi f G m)^{-\frac23} \\
&+ \left[ T F_b + Q e^{-i\varphi_{(2,0)}} P_{(2,0)} \right] \\
&+ \left[ e^{-i\varphi_{(2,1)}}P_{(2,1)}+e^{-i\varphi_{(2,0)}}P_{(2,0)}S_1 \right](\pi f G m)^{\frac23}  \\
&+ \left[ e^{-i\varphi_{(2,3/2)}}P_{(2,3/2)}+e^{-i\varphi_{(2,0)}}P_{(2,0)}S_{3/2} \right](\pi f G m)  \\
&+ \left[ e^{-i\varphi_{(2,2)}}P_{(2,2)}+e^{-i\varphi_{(2,1)}}P_{(2,1)}S_{1}+e^{-i\varphi_{(2,0)}}P_{(2,0)}S_2 \right](\pi f G m)^{\frac43}  \\
&+ \left[ e^{-i\varphi_{(2,5/2)}}P_{(2,5/2)}+e^{-i\varphi_{(2,3/2)}}P_{(2,3/2)}S_{1}+e^{-i\varphi_{(2,1)}}P_{(2,1)}S_{3/2}+e^{-i\varphi_{(2,0)}}P_{(2,0)}S_{5/2} \right](\pi f G m)^{\frac53}
\Bigr\}  \\
&\times \Theta(2f_{\rm LSO}-f) \exp
\left\{
i \left[ 2\pi ft_c-\frac{\pi}{4}+2\Psi(f/2) \right]
\right\},
\end{aligned}
\end{equation}

\begin{equation}
\begin{aligned}
\tilde{h}^{(3)}(f) ={} &3^{-\frac12}\left(\frac{5}{48}\right)^{\frac12}\pi^{-\frac23}\frac{(G M_c)^{5/6}}{D}(2f/3)^{-\frac76}
\Bigl\{
e^{-i\varphi_{(3,1/2)}}P_{(3,1/2)}(2\pi f G m/3)^{\frac13}  \\
&+ \left [e^{-i\varphi_{(3,3/2)}}P_{(3,3/2)}+e^{-i\varphi_{(3,1/2)}}P_{(3,1/2)}S_1 \right](2\pi f G m/3)  \\
&+ \left[ e^{-i\varphi_{(3,2)}}P_{(3,2)}+e^{-i\varphi_{(3,1/2)}}P_{(3,1/2)}S_{3/2} \right](2\pi f G m/3)^{\frac43}  \\
&+ \left[ e^{-i\varphi_{(3,3/2)}}P_{(3,3/2)}S_1+e^{-i\varphi_{(3,1/2)}}P_{(3,1/2)}S_{2} \right](2\pi f G m/3)^{\frac53}
\Bigr\}  \\
&\times \Theta(3f_{\rm LSO}-f)\exp
\left\{
i \left[ 2\pi ft_c-\frac{\pi}{4}+3\Psi(f/3) \right]
\right\},  \\
\end{aligned}
\end{equation}

\begin{equation}
\begin{aligned}
\tilde{h}^{(4)}(f) ={} &4^{-\frac12}\left(\frac{5}{48}\right)^{\frac12}\pi^{-\frac23}\frac{(G M_c)^{5/6}}{D}(f/2)^{-\frac76}
\Bigl\{
e^{-i\varphi_{(4,1)}}P_{(4,1)}(\pi f G m/2)^{\frac23}  \\
&+ \left[ e^{-i\varphi_{(4,2)}}P_{(4,2)}+e^{-i\varphi_{(4,1)}}P_{(4,1)}S_1 \right](\pi f G m/2)^{\frac43}  \\
&+ \left[ e^{-i\varphi_{(4,5/2)}}P_{(4,5/2)}+e^{-i\varphi_{(4,1)}}P_{(4,1)}S_{3/2} \right](\pi f G m/2)^{\frac53}
\Bigr\}  \\
&\times \Theta(4f_{\rm LSO}-f)\exp
\left\{
i \left[ 2\pi ft_c-\frac{\pi}{4}+4\Psi(f/4) \right]
\right\},  \\
\end{aligned}
\end{equation}

\begin{equation}
\begin{aligned}
\tilde{h}^{(5)}(f) ={} &5^{-\frac12}\left(\frac{5}{48}\right)^{\frac12}\pi^{-\frac23}\frac{(G M_c)^{5/6}}{D}(2f/5)^{-\frac76}
\Bigl\{
e^{-i\varphi_{(5,3/2)}}P_{(5,3/2)}(2\pi f G m/5)  \\
&+ \left[ e^{-i\varphi_{(5,5/2)}}P_{(5,5/2)}+e^{-i\varphi_{(5,3/2)}}P_{(5,3/2)}S_{1} \right](2\pi f G m/5)^{\frac53}
\Bigr\}  \\
&\times \Theta(5f_{\rm LSO}-f)\exp
\left\{
i \left[ 2\pi ft_c-\frac{\pi}{4}+5\Psi(f/5) \right]
\right\},  \\
\end{aligned}
\end{equation}

\begin{equation}
\begin{aligned}
\tilde{h}^{(6)}(f) ={} &6^{-\frac12}\left(\frac{5}{48}\right)^{\frac12}\pi^{-\frac23}\frac{(G M_c)^{5/6}}{D}(f/3)^{-\frac76}
\Bigl\{
e^{-i\varphi_{(6,2)}}P_{(6,2)}(\pi f G m/3)^{\frac43}
\Bigr\} \\
&\times \Theta(6f_{\rm LSO}-f)\exp
\left\{
i \left[ 2\pi ft_c-\frac{\pi}{4}+6\Psi(f/6) \right]
\right\},  \\
\end{aligned}
\end{equation}

\begin{equation}
\begin{aligned}
\tilde{h}^{(7)}(f) ={} &7^{-\frac12}\left(\frac{5}{48}\right)^{\frac12}\pi^{-\frac23}\frac{(G M_c)^{5/6}}{D}(2f/7)^{-\frac76}
\Bigl\{
e^{-i\varphi_{(7,5/2)}}P_{(7,5/2)}(2\pi f G m/7)^{\frac53}
\Bigr\} \\
&\times \Theta(7f_{\rm LSO}-f)\exp
\left\{
i \left[ 2\pi ft_c-\frac{\pi}{4}+7\Psi(f/7) \right]
\right\},
\end{aligned}
\end{equation}
where
\begin{equation}
\begin{aligned}
S_1 &= \frac{1}{2} \left( \frac{743}{336}+\frac{11}{4}\eta \right), \\
S_{3/2} &= -2\pi, \\
S_2 &= \frac{7266251}{8128512}+\frac{18913}{16128}\eta+\frac{1379}{1152}\eta^2,  \\
S_{5/2} &=-\pi\frac{4757}{1344}-\frac{3}{16}(-63+44\pi)\eta,
\end{aligned}
\end{equation}
and
\begin{equation}
e^{-i\varphi_{(n,s)}}P_{(n,s)} =\left[F_+ C_+^{(n,s)} + F_\times C_\times^{(n,s)}\right] + i \left[F_+ D_+^{(n,s)} + F_\times D_\times^{(n,s)}\right].
\end{equation}
$C_{+,\times}^{(n,s)}$ and $D_{+,\times}^{(n,s)}$ denote the prefactors of $\cos(n\Psi)$ $\sin(n\Psi)$ in $H_{+,\times}^{(s)}$ respectively, which can be found in references \citep{Arun_2004, Arun_2005}, and we will not repeat them here.
$f_{\rm LSO}$ is the last orbital frequency where the waveforms are truncated. We employ the Roche radius of rigid spherical bodies in Eq. (\ref{roche_r}) as a rough estimation of last stable distance between two objects of a binary.
{The SMG corrections enter the waveform in the harmonic one (\ref{h1}) and the harmonic two (\ref{h2}). The terms relevant to $(2\pi f G m)^{-1}$ and $(2\pi f G m)^{-\frac13}$ in the harmonic one (\ref{h1}) are directly added into the waveform. These two terms are corresponding to the the Eq.(\ref{hb_1}). 
The term relevant to $(\pi f G m)^{-\frac23}$ in the harmonic two (\ref{h2}), which is corresponding to the terms having the same power of $(\pi f G m)$ in Eq.(\ref{hb_2}) and Eq.(\ref{h_p_c}), is also directly added into the waveform.
The above three terms are 0 in the waveform of GR.
The term relevant to $(\pi f G m)^0$ in the harmonic two (\ref{h2}) is the modified leading order term which can return to the case of GR when $T F_b=0$ and $Q=1$.
The rest terms are all from amplitude higher order corrections in GR which can also be found in the reference \citep{van2006phenomenology}.}
Different from ground-based detectors, the antenna pattern functions of space-borne detectors depend on time. In the Fourier transform of a detector's response function got by using stationary phase approximation, the time $t$ in $F_+$, $F_\times$ and $F_b$ are replaced by function $t(f)$ which are given by
\bq
t(f) = t_c-\frac{5}{256(G M_c)^{5/3}}(2\pi f)^{-8/3} \sum_{i=-2}^{7} \tau_i (2\pi f G m)^{i/3},
\eq
with the coefficients
\begin{equation}
\begin{aligned}
\tau_{-2} ={} &-\frac{1}{48}\epsilon_d^2,  \\
\tau_{-1} ={} &0,  \\
\tau_0 ={} &1,  \\
\tau_1 ={} &0,  \\
\tau_2 ={} &\frac{743}{252} + \frac{11}{3}\eta,  \\
\tau_3 ={} &-\frac{32}{5}\pi,  \\
\tau_4 ={} &\frac{3058673}{508032} + \frac{5429}{504}\eta + \frac{617}{72}\eta^2,  \\
\tau_5 ={} &-\left( \frac{7729}{252}-\frac{13}{3}\eta \right)\pi,  \\
\tau_6 ={} &-\frac{10052469856691}{23471078400} + \frac{128\pi^2}{3} + \frac{6848\gamma}{105} +\left(\frac{3147553127}{3048192} - \frac{451\pi^2}{12} \right)\eta  \\
&-\frac{15211}{1728}\eta^2 + \frac{25565}{1296}\eta^3 +\frac{3424}{105}\ln[16(2\pi m f)^{2/3}],  \\
\tau_7 ={} &\left(-\frac{15419335}{127008} - \frac{75703}{756}\eta + \frac{14809}{378}\eta^2 \right)\pi.
\end{aligned}
\end{equation}
Here, $\gamma=0.5772$ is the Euler-Mascheroni constant, and $\tau_{-2}$ is concerned with the corrections induced by SMG, $\tau_{i}(i\ge0)$ is concerned with the frequency evolution at 3.5 PN in phase \citep{buonanno2009comparison}.
The phase $\Psi(f)$ is given by
\begin{equation} \label{phase}
\Psi(f) = -\Psi_c + \frac{3}{ 256 (2\pi f G M_c)^{5/3}} \sum_{i=-2}^{7} \Psi_i (2\pi f G m)^{i/3} - \Psi_D[t(f)],
\end{equation}
where the coefficients $\Psi_i$ are given by
\begin{equation}
\begin{aligned}
\Psi_{-2} ={} &-\frac{5}{336}\epsilon_d^2,  \\
\Psi_{-1} ={} &0,  \\
\Psi_0 ={} &1,  \\
\Psi_1 ={} &0,  \\
\Psi_2 ={} &\frac{20}{9}\,\left[\frac{743}{336} + \frac{11}{4}\eta\right],\nn\\
\Psi_3 ={} &-16\pi,  \\
\Psi_4 ={} &10\,\left[\frac{3058673}{1016064} + \frac{5429}{1008}\eta + \frac{617}{144}\eta^2 \right],  \\
\Psi_5 ={} &\pi\,\left[\frac{38645}{756} + \frac{38645}{756}\ln\left(\frac{f}{f_{LSO}}\right) - \frac{65}{9}\eta\left(1 + \ln\left(\frac{f}{f_{LSO}}\right)\right)\right],  \\
\Psi_6 ={} &\left(\frac{11583231236531}{4694215680} - \frac{640\pi^2}{3} - \frac{6848\gamma}{21}\right)  \\
& + \eta\,\left(-\frac{15737765635}{3048192} + \frac{2255\pi^2}{12}\right)  \\
& + \frac{76055}{1728}\eta^2 - \frac{127825}{1296}\eta^3 - \frac{6848}{21}\ln\left[4(2\pi f G m)^{1/3}\right],  \\
\Psi_7 ={} &\pi\left(\frac{77096675}{254016} + \frac{378515}{1512}\eta-\frac{74045}{756}\eta^2\right),  \\
\label{psi_coefficients}
\end{aligned}
\end{equation}
with $\gamma=0.5772$ the Euler-Mascheroni constant.
$\Psi_i(i\ge0)$ is the coefficients in 3.5PN phase function of Fourier domain waveform, and the coefficient $\Psi_{-2}$ is concerned with the correction of dipole radiation in SMG.
$\Psi_D$ denotes the Doppler modulation which is the difference between the phase of the wavefront at the detector and the barycenter. The expression of $\Psi_D$ is given by \citep{cutler1998angular,hu2018fundamentals},
\bq
\Psi_D = 2 \pi f R \sin\theta \cos\left[\frac{2\pi t(f)}{T} + b_0 -\varphi\right],
\eq
where $\theta$ and $\varphi$ are the ecliptic colatitude and longitude of the GW source, $R=1 \rm{A.U.}$, $T$ is one year, and $b_0$ is the ecliptic longitude of the detector at $t=0$.
Finally, the other parameters, $E, Q, S_{-1}, T$, are defined by Eqs. \ref{def_E}, \ref{def_Q}, \ref{def_S-1} and \ref{def_T}.

\section{antenna pattern function} \label{app_pattern_f}
The response functions of LISA-like detectors can be found in the references \citep{cornish2003lisa, rubbo2004forward, cornish2001space, PhysRevD.99.104027}. In this Appendix, we will give the expressions of antenna pattern function in a specific coordinate system.

The general form of antenna pattern function for LISA-like detectors is given in Eqs. \ref{pattern_f} and \ref{transfer_f}.
In the low frequency range, the transfer functions ${\mathcal T}(f, \hat{\boldsymbol l}_1\cdot \hat{\boldsymbol \Omega})$ and ${\mathcal T}(f, \hat{\boldsymbol l}_2\cdot \hat{\boldsymbol \Omega})$ approach to 1, and the antenna pattern functions return to the cases which are similar with the ground-based detectors.
Here we will write explicitly the polarization tensor $\epsilon_{ij}^A$ and the vectors $\hat{\boldsymbol l}_1$, $\hat{\boldsymbol l}_2$, $\hat{\boldsymbol \Omega}$ in a specific coordinate system. This process is similar with the case of ground-based detectors \citep{maggiore2008gravitational, poisson2014gravity} or the case of space-borne detectors in low frequency approximation \citep{cutler1998angular, hu2018fundamentals}.

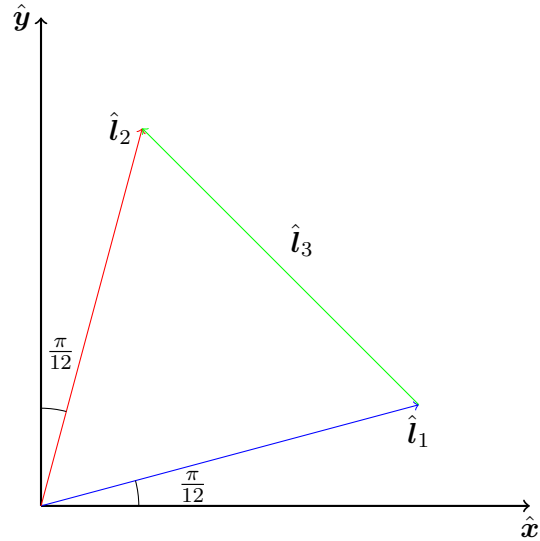
\begin{figure}
\centering
	\begin{tikzpicture}[scale=1.3]
	\draw[->, thick] (0,0) -- (5,0);
	\node [below] at (5,0) {$\hat{\boldsymbol x}$};
	\draw[->, thick] (0,0) -- (0,5);
	\node [left] at (0,5) {$\hat{\boldsymbol y}$};
	
	\draw[->, color = blue] (0,0) -- (3.8637,1.0353);
	\node [below] at (3.8637,1.0353) {$\hat{\boldsymbol l}_1$};
	\draw[->, color = red] (0,0) -- (1.0353,3.8637);
	\node [left] at (1.0353,3.8637) {$\hat{\boldsymbol l}_2$};
	\draw[->, color = green] (3.8637,1.0353) -- (1.0353,3.8637);
	\node [above right] at (2.4495,2.4495) {$\hat{\boldsymbol l}_3$};
	
	\draw (75:1) arc (75:90:1);
	\node [above] at (0.2,1.3) {$\frac{\pi}{12}$};
	\draw (0:1) arc (0:15:1);
	\node [right] at (1.3,0.2) {$\frac{\pi}{12}$};
	\end{tikzpicture}
	\caption{the detector coordinate}
\label{fig_dete_coord}
\end{figure}

We choose the coordinate system tied with the detector, which is denoted by $\hat{\boldsymbol x}\hat{\boldsymbol y}\hat{\boldsymbol z}$. The interferometer arms are put in this coordinate as shown in Fig. \ref{fig_dete_coord}. The unit vectors of two arms can be expressed as
\bq \label{arms}
\begin{aligned}
  \hat{\boldsymbol l}_1 &=
	\begin{pmatrix}
	\cos\frac{\pi}{12} \\
	\sin\frac{\pi}{12} \\
	0
	\end{pmatrix},
~~~~
  \hat{\boldsymbol l}_1 &=
	\begin{pmatrix}
	\cos\frac{5\pi}{12} \\
	\sin\frac{5\pi}{12} \\
	0
	\end{pmatrix},
\end{aligned}
\eq
in this coordinate.

In a general metric theory, there are up to six possible polarization modes. Besides the $h_+$ and $h_\times$ modes in GR, there are purely transverse $h_b$ mode, purely longitudinal $h_l$ mode, and two mixed modes $h_x$ and $h_y$. In the coordinate $(\hat{\boldsymbol x}_1, \hat{\boldsymbol x}_2, \hat{\boldsymbol x}_3)$ where the GW travels along $\hat{\boldsymbol x}_3$ direction, these polarizations can be expressed as
\bq
h_{ij} =
	\begin{pmatrix}
	h_b+h_+	&h_\times	&h_{x} \\
	h_\times	&h_b-h_+		&h_{y} \\
	h_{x}		&h_{y}			&h_l
	\end{pmatrix}.
\eq
In SMG, there are four modes $h_+$, $h_\times$, $h_b$ and $h_l$ \citep{liu2018waveforms}. We can use polarization tensors $\epsilon^A_{ij}$ to expand the metric perturbation as
\bq
h_{ij}(t) = \sum_{A} \epsilon_{ij}^A h_A(t),
\eq
where $A = +, \times, b, l$ labels the polarization modes. By using the unit vector $\hat{\boldsymbol \Omega}$ (which points the propagation direction of the GW), and the unit vectors $\hat{\boldsymbol u}$ and $\hat{\boldsymbol v}$ (which are orthogonal to $\hat{\boldsymbol \Omega}$ and orthogonal to each other), the polarization tensors can be rewritten as
\bq \label{pol_tensor}
\begin{aligned}
\epsilon^+_{ij} 				&= \hat{u}_i\hat{u}_j - \hat{v}_i\hat{v}_j, ~~~~
\epsilon^{\times}_{ij} 	&= \hat{u}_i\hat{v}_j + \hat{v}_i\hat{u}_j, ~~~~
\epsilon^b_{ij} 				&= \hat{u}_i\hat{u}_j + \hat{v}_i\hat{v}_j, ~~~~
\epsilon^l_{ij} 				&= \hat{\Omega}_i\hat{\Omega}_j.
\end{aligned}
\eq
Thus, the $F_A$ can be derived straightforwardly, once the unit vectors $\hat{\boldsymbol \Omega}$, $\hat{\boldsymbol u}$, $\hat{\boldsymbol v}$ and $\hat{\boldsymbol l}_1$, $\hat{\boldsymbol l}_2$ are expressed in the same coordinate.
We employ $(\theta', \varphi')$ to represent the direction of the GW source in the detector coordinate, where $\varphi'$ is the azimuth angle and $\theta'$ is the altitude angle (we use the definition of $\theta'$ which is the angle between the direction of GW source and the direction of $\hat{\boldsymbol z}$ in this work). And $\psi'$ denotes the polarization angle in the detector coordinate. Therefore, the vectors $\hat{\boldsymbol \Omega}$, $\hat{\boldsymbol u}$, $\hat{\boldsymbol v}$ in the detector coordinate can be given by
\bq \label{uvO2ijk}
\begin{aligned}
\hat{\boldsymbol u} &=
	\begin{pmatrix}
	\cos\theta' \cos\varphi' \cos\psi' - \sin\varphi' \sin\psi' \\
	\cos\theta' \sin\varphi' \cos\psi' + \cos\varphi' \sin\psi' \\
	-\sin\theta' \cos\psi'
	\end{pmatrix},
\\
\hat{\boldsymbol v} &=
	\begin{pmatrix}
	\cos\theta' \cos\varphi' \sin\psi' + \sin\varphi' \cos\psi' \\
	\cos\theta' \sin\varphi' \sin\psi' - \cos\varphi' \cos\psi' \\
	-\sin\theta' \sin\psi'
	\end{pmatrix},
\\
\hat{\boldsymbol \Omega} &=
	\begin{pmatrix}
	- \sin\theta' \cos\varphi' \\
	- \sin\theta' \sin\varphi' \\
	-\cos\theta'
	\end{pmatrix}.
\end{aligned}
\eq
Different from the ground-based detectors, the time scale of GW signal detected by space-borne detectors is comparable with the time scale of detectors' motion. The motion of space-borne detectors cannot be neglected, so $(\theta', \varphi', \psi')$ are considered to be time-dependent.
We need to find the relationships between the direction of GW source as well as the polarization angle in the detector coordinate and those in the heliocentric coordinate which are denoted by $(\theta, \varphi, \psi)$.
The relationships are depending on the motion of detectors, We will discuss LISA first.


We employ $\hat{\boldsymbol i}\hat{\boldsymbol j}\hat{\boldsymbol k}$ to denote the heliocentric coordinate tied with the ecliptic, and $(\theta, \varphi, \psi)$ to represent the direction of GW source and the polarization angel in the heliocentric coordinate respectively.
Recalling the orbital configuration of LISA, which have been introduced in Section \ref{sec_detectors},
the unit vectors $\hat{\boldsymbol x}$, $\hat{\boldsymbol y}$ and $\hat{\boldsymbol z}$ of the detector coordinate can be written in terms of the heliocentric coordinate as
\bq
\begin{aligned}
\hat{\boldsymbol x} &= \left[\frac{1}{2}\cos a(t) \cos b(t) + \sin a(t) \sin b(t)\right]\hat{\boldsymbol i}
+\left[\frac{1}{2}\cos a(t) \sin b(t) - \sin a(t) \cos b(t) \right]\hat{\boldsymbol j}
+\left[\frac{\sqrt{3}}{2}\cos a(t) \right]\hat{\boldsymbol k}
\\
\hat{\boldsymbol y} &= \left[\frac{1}{2}\sin a(t) \cos b(t) - \cos a(t) \sin b(t) \right]\hat{\boldsymbol i}
+\left[\frac{1}{2}\sin a(t) \sin b(t) + \cos a(t) \cos b(t) \right]\hat{\boldsymbol j}
+\left[\frac{\sqrt{3}}{2}\sin a(t) \right]\hat{\boldsymbol k}
\\
\hat{\boldsymbol z} &= \left[-\frac{\sqrt{3}}{2}\cos b(t) \right] \hat{\boldsymbol i}
+ \left[-\frac{\sqrt{3}}{2}\sin b(t) \right] \hat{\boldsymbol j}
+\left(\frac{1}{2}\right)\hat{\boldsymbol k},
\end{aligned}
\eq
where $a(t) = a_0 + \frac{2\pi t}{T_{\rm LISA}}$ is the phase of rotation around detector's center, and $b(t) =b_0 + \frac{2\pi t}{T_{\rm LISA}}$ is the phase of revolution around the sun. For the motion of LISA, The periods of rotation around the detector's center and the revolution around the sun are both one year. The initial phase $a_0$ and $b_0$ are constant. We can take $a_0 = 0$ and $b_0 = 0$ without loss of generality. And the direction of GW source $\hat{\boldsymbol r}$ can be given in $\hat{\boldsymbol i}\hat{\boldsymbol j}\hat{\boldsymbol k}$ coordinate as
\bq
\hat{\boldsymbol r} = (\sin\theta\cos\varphi) \hat{\boldsymbol i}
+(\sin\theta\sin\varphi) \hat{\boldsymbol j}
+(\cos\theta) \hat{\boldsymbol k}.
\eq
Using the geometry relationships of those vectors and angles, $\theta'$ and $\varphi'$ can be get by
\bq
\cos\theta' = \hat{\boldsymbol r} \cdot \hat{\boldsymbol z},
~~~
\tan\phi' = \frac{\hat{\boldsymbol r} \cdot \hat{\boldsymbol y}}{\hat{\boldsymbol r} \cdot \hat{\boldsymbol x}}.
\eq
As for polarization angle $\psi'$, we follow the definition in the reference \citep{cutler1998angular}.
An ellipse can be got by projecting the binary's circular orbit on the plane of the sky (i.e., the plane orthogonal to the GWs' propagation direction). The major axis of this ellipse are defined as the vector $\hat{\boldsymbol u}$ mentioned above. The polarization angle is defined as the angel between the vector $\hat{\boldsymbol u}$ and the vector pointing to the direction increasing $\theta'$. A useful figure can be referred to in the literature \citep{poisson2014gravity} (figure 11.5). According to this definition, the polarization is given by
\bq
\tan\psi' = \frac{\bigl[ \hat{\boldsymbol L}-(\hat{\boldsymbol L}\cdot\hat{\boldsymbol r})\hat{\boldsymbol r} \bigr] \cdot \hat{\boldsymbol z}}{(\hat{\boldsymbol r}\times\hat{\boldsymbol L}) \cdot \hat{\boldsymbol z}},
\eq
where $\hat{\boldsymbol L}$ denotes the unit vector parallel to orbital angular momentum vector of the binary.
The vector $\hat{\boldsymbol L}$ in the detector coordinate $\hat{\boldsymbol x}\hat{\boldsymbol y}\hat{\boldsymbol z}$ is time-dependent, therefore we prefer to express $\hat{\boldsymbol L}$ in the heliocentric coordinate $\hat{\boldsymbol i}\hat{\boldsymbol j}\hat{\boldsymbol k}$.
The vector $\hat{\boldsymbol L}$ in the coordinate $\hat{\boldsymbol i}\hat{\boldsymbol j}\hat{\boldsymbol k}$ can be given by
\bq
\begin{aligned}
\hat{\boldsymbol L} ={}
&\Bigl[\cos\iota\sin\theta\cos\varphi
+\sin\iota\bigl(\cos\psi\sin\varphi+\cos\theta\cos\varphi\sin\psi\bigr)\Bigr]\hat{\boldsymbol i} \\
&+\Bigl[-\sin\iota\cos\psi\cos\varphi
+\sin\varphi\bigl(\cos\iota\sin\theta+\cos\theta\sin\iota\sin\psi\Bigr)\bigr]\hat{\boldsymbol j} \\
&+\Bigl[\cos\iota\cos\theta - \sin\theta\sin\iota\sin\psi\Bigr]\hat{\boldsymbol k}.
\end{aligned}
\eq
In the above equation, inclination angle $\iota$ is the angle between $\hat{\boldsymbol L}$ and $\hat{\boldsymbol r}$, the polarization angle $\psi$ in heliocentric coordinate $\hat{\boldsymbol i}\hat{\boldsymbol j}\hat{\boldsymbol k}$ has the similar definition with $\psi'$ in the coordinate $\hat{\boldsymbol x}\hat{\boldsymbol y}\hat{\boldsymbol z}$, which is the angle between the major axis of the projection ellipse and the vector pointing to the direction increasing $\theta$.

The parameters $(\theta', \varphi', \psi')$ in the detector coordinate can be eventually expressed in terms of the parameters $(\theta, \varphi, \psi, \iota)$ in the heliocentric coordinate, which are considered to be time-independent and the variables $a(t)$, $b(t)$ describing the motion of detectors in the heliocentric coordinate which have simple relationships with time.
Substituting  Eq. \ref{uvO2ijk} into Eq. \ref{pol_tensor}, we can obtain the polarization tensors in the detector coordinate. The transfer functions can also be obtained by taking the expressions of $\hat{\boldsymbol l}_1$, $\hat{\boldsymbol l}_2$, $\hat{\boldsymbol \Omega}$ (Eqs. \ref{uvO2ijk}, \ref{arms}) into the Eq. \ref{transfer_f}. Substituting these results and Eq. \ref{arms} into Eq. \ref{pattern_f}, the antenna pattern functions can be assembled finally. The final results are straightforward but cumbersome. To avoid redundancy, the results are not presented here.

The similar calculation is also applicable for TianQin. Comparing with LISA, TianQin will run in a geocentric orbit, and the orientation of TianQin is fixed to the reference source J0806.3+1527 instead of varying with time. A brief introduction of TianQin's orbit have been given in Section \ref{sec_detectors}.
The base vectors of the TianQin's detector coordinate $(\hat{\boldsymbol x}, \hat{\boldsymbol y},\hat{\boldsymbol z})_{\rm TianQin}$ can be given by
\bq
\begin{aligned}
\hat{\boldsymbol x} &= \Bigl[\cos a(t) \cos\theta_0\cos\varphi_0 - \sin a(t) \sin\varphi_0\Bigr]\hat{\boldsymbol i}
+\Bigl[\cos a(t) \cos\theta_0\sin\varphi_0 + \sin a(t) \cos\varphi_0\Bigr]\hat{\boldsymbol j}
+\Bigl[-\cos a(t) \sin\theta_0\Bigr]\hat{\boldsymbol k},
\\
\hat{\boldsymbol y} &= \Bigl[-\sin a(t) \cos\theta_0\cos\varphi_0 - \cos a(t) \sin\varphi_0\Bigr]\hat{\boldsymbol i}
+\Bigl[-\sin a(t) \cos\theta_0\sin\varphi_0 + \cos a(t) \cos\varphi_0\Bigr]\hat{\boldsymbol j}
+\Bigl[\sin a(t) \sin\theta_0\Bigr]\hat{\boldsymbol k},
\\
\hat{\boldsymbol z} &= \Bigl(\sin\theta_0\cos\varphi_0\Bigr) \hat{\boldsymbol i}
+\Bigl(\sin\theta_0\sin\varphi_0\Bigr) \hat{\boldsymbol j}
+\Bigl(\cos\theta_0\Bigr) \hat{\boldsymbol k},
\end{aligned}
\eq
in the heliocentric coordinate, where $(\theta_0, \varphi_0)$ denote the direction of the reference source, and $a(t) = a_0 + \frac{2\pi t}{T_{\rm TianQin}}$ represent the rotation phase of the detector.
$T_{\rm TianQin}$ is the period of TianQin's rotation which is about 3.65 days.
Following the same process discussed above, the antenna pattern functions of TianQin can be derived.

\bibliography{ref}

\end{document}